\documentclass[preprint2]{aastex}

\shortauthors{Turpin et al.}

\usepackage{graphicx}
\usepackage{txfonts}
\usepackage{float}
\usepackage{natbib}
\bibpunct{(}{)}{;}{a}{}{,} 

\def\epi{E$_{\rm pi}$}

\def\eiso{E$_{\rm iso}$}
\def\liso{L$_{\rm iso}$}
\def\eer{\epi\ -- \eiso\ {\rm relation}}
\def\eep{\epi\ -- \eiso\ {\rm plane}}

\def\elr{\epi\ -- \liso\ {\rm relation}}
\def\elp{\epi\ -- \liso\ {\rm plane}}

\def\t9{T$_{90}$}

\def\nallswift{90}
\def\nall{126}

\def\ngrb{76}
\def\ngrbnozall{42}
\def\ngrbnoz{14}

\begin{document}


\title{Investigating the impact of optical selection effects on observed rest frame prompt GRB properties}


\author{D. Turpin\altaffilmark{1,2}, V. Heussaff\altaffilmark{1,2}, J.-P. Dezalay\altaffilmark{1,2}, J-L. Atteia\altaffilmark{1,2}, A. Klotz\altaffilmark{1,2} and D. Dornic\altaffilmark{3}} 

\email{damien.turpin@irap.omp.eu}


\altaffiltext{1}{Universit\'e de Toulouse; UPS-OMP; IRAP; Toulouse, France}
\altaffiltext{2}{CNRS; IRAP; 14, avenue Edouard Belin, F-31400 Toulouse, France}
\altaffiltext{3}{Aix Marseille Universit\'e, CNRS/IN2P3, CPPM UMR 7346, 13288 Marseille, France}


\begin{abstract}
Measuring gamma-ray burst (GRB) properties in their rest frame is crucial for understanding the physics at work in GRBs. This can only be done for GRBs with known redshifts. Since redshifts are usually measured from the optical spectrum of the afterglow, correlations between prompt and afterglow emissions may introduce biases into the distribution of the rest-frame properties of the prompt emission, especially considering that we measure the redshift of only one-third of {\it Swift} GRBs. 
   In this paper, we study the optical flux of GRB afterglows and its connection to various intrinsic properties of GRBs. We also discuss the impact of the optical selection effect on the distribution of rest-frame prompt properties of GRBs. 
   
   
   Our analysis is based on a sample of \nallswift\ GRBs with good optical follow-up and well-measured prompt emission. Seventy-six of them have a measure of redshift and 14 have no redshift.
   We compare the rest-frame prompt properties of GRBs with different afterglow optical fluxes in order to check for possible correlations between the promt properties and the optical flux of the afterglow. The optical flux is measured two hours after the trigger, which is a typical time for the measure of the redshift.
   
   
We find that the optical flux of GRB afterglows in our sample is mainly driven by their optical luminosity and depends only slightly on their redshift. We show that GRBs with low and high afterglow optical fluxes have similar \epi, \eiso, and \liso, indicating that the rest-frame distributions computed from GRBs with a redshift are not significantly distorted by optical selection effects. However, we found that the $T_{90}^{rest}$ distribution is not immune to optical selection effects, which favor the selection of GRBs with longer durations. Finally, we note that GRBs well above the \eer\ have lower optical fluxes and we show that optical selection effects favor the detection of GRBs with bright optical afterglows located close to or below the best-fit \eer\ (Amati relation), whose redshift is easily measurable.

   With more than 300 GRBs with a redshift, we now have a much better view of the intrinsic properties of these remarkable events. At the same time, increasing statistics allow us to understand the biases acting on the measurements. The optical selection effects
induced by the redshift measurement strategies cannot be neglected when we study the properties of GRBs in their rest-frame, even for studies focused on the prompt emission.
   
\end{abstract}



   \maketitle


\section{Introduction}
\label{sec_intro}

Gamma-ray bursts (GRBs) are cataclysmic explosions resulting from the collapse of massive stars or the merging of two compact objects (see \cite{Kumar2014} and references therein). During these events, an ultra relativistic jet is produced that is accompanied by an intense gamma-ray flash which can be seen at very high redshifts (up to $z=9$ for GRB090429B; see \cite{Cucchiara2011a}. The gamma-ray emission is produced in a few seconds (prompt emission) with a total energy release of up to E$\sim10^{54}$ erg assuming an isotropic emission. This prompt emission is followed by a long-lasting and fading multi-wavelength emission from X-rays to radio (afterglow emission) which can be attributed to the external shock between the relativistic ejecta and the interstellar medium (ISM). The prompt $\gamma$-ray emission has been suggested as being due to internal shocks between shells moving at different Lorentz factors, $\Gamma$, inside the jet (see \citep[][]{Rees1994,Paczynski1994,Piran1999,Kumar2014}). However, the physical processes at work in these shocks are still not well understood due to strong uncertainties on the physical conditions (baryon loading, energy dissipation in the jet, acceleration mechanism; see the review by \cite{Kumar2014}). On the contrary, the afterglow emission is better understood and can be explained by the synchrotron emission from accelerated electrons in the shock between the relativistic ejecta and the ISM (forward shock; (see \cite{Sari1998} $\&$ \cite{Granot2002}). An optical/radio flash is also expected and is sometimes observed from a reverse shock propagating into the relativistic ejecta (see \citep[][]{Meszaros1993, Kobayashi2000}). Studying the optical flash radiation from the reverse shock provides us with the opportunity to constrain the magnetization parameter of the ejecta \citep[][]{Zhang2005, Narayan2011}, the Lorentz factor of the jet, and also on the jet composition \citep[][]{McMahon2006, Nakar2004}. In this framework, the dynamics of the afterglow is determined by the microphysics of the shocked material, the ISM properties, and the kinetic energy of the blast wave, $E_k$, which depends on the prompt emission properties (the isotropic gamma-ray energy released, $E_{iso}$, and the gamma-ray radiative efficiency, $\eta$). Understanding the connection between the prompt and afterglow properties would help us to better constrain the physics of GRBs and relativistic jets. 

\smallskip 

Many authors have discussed correlations between the afterglow luminosity and the prompt GRB energetics. Correlations between the afterglow optical luminosity and prompt isotropic energy have been found by \cite{Kann2010} and \cite{Nysewander2009}, and between the afterglow X-ray emission and the isotropic energy by \cite{Kaneko2007} and \cite{Margutti2013}. However, it is difficult to assess whether or not these relations have their origin in the physics of the GRB because some studies have shown that they could result from selection effects. Indeed, \cite{Coward2014} recently detected a strong Malmquist bias in the correlation $E_{iso}-L_{opt,X}$ as we preferentialy detect the brightest GRBs. While it is clear that gamma-ray selection effects can bias statistical studies of prompt GRB properties, the impact of optical selection effects is rarely assessed. This paper is dedicated to the study of potential selection effects in the distribution of rest-frame prompt properties due to the need of measuring the redshift.
We construct a sample of GRBs with good optical and gamma-ray data in section \ref{sec_sample}. The potential parameters (extrinsic and intrinsic) that could bias our sample are also investigated in section \ref{sec_analyse_opt}. The main conclusion of this section is that the distribution of the afterglow optical flux is mainly driven by their optical luminosities as opposed to redshift distribution. Thus, in section \ref{sec:sel_effect}, we show how the need to detect the optical afterglow can act against the measurement of the rest-frame properties of some GRBs. Then, in section \ref{sec_prompt} we compare the afterglow optical flux of GRBs with their typical rest-frame prompt properties as $E_{iso}$ or $L_{iso}$ to estimate the impact of the optical selection effect. We also investigate optical selection effects on GRB rest-frame correlations in section \ref{sec_correlation}. In section \ref{discussion}, we briefly discuss the consequences of our findings.


\section{GRB sample and data}
\label{sec_sample}

\subsection{Gamma-ray data}
\label{sub_HE}

Our gamma-ray selection criteria followed a procedure similar to that of \cite{Heussaff2013}.
We selected a large sample of GRBs with well-measured spectral parameters available in the literature, i.e GCN Circulars\footnote{http://gcn.gsfc.nasa.gov}. This covers GRB observations from 1997 to 2014. GRB spectra are parametrized with the Band function, \cite{Band1993}, which consists of two smoothly connected power laws :

\begin{equation}
\begin{array}{clllll}
F_\gamma(E_\gamma) =\\
\\
 f_\gamma \left\lbrace
\begin{array}{ll}
(\frac{E_\gamma}{\rm{100~keV}})^{\alpha_\gamma} ~ \rm{exp}(-\frac{E_\gamma(\alpha_\gamma-\beta_\gamma)}{E_b}), ~ if~E_\gamma < E_b\\
\\
(\frac{E_b}{\rm{100~keV}})^{\alpha_\gamma-\beta_\gamma} ~ \rm{exp}(\beta_\gamma-\alpha_\gamma)~ (\frac{E_\gamma}{\rm{100~keV}})^{\beta_\gamma},\\
\\
\rm{ if~E_\gamma \ge E_b}\\
\end{array}
\right.
\end{array}\\
\end{equation}
where $E_b = \frac{\alpha_\gamma-\beta_\gamma}{2+\alpha_\gamma}\times E_{po}$. $\alpha$ ($\alpha >-2$) and $\beta$ ($\beta < -2$) are the low- and high-energy photon spectral indexes of the power law, respectively. $E_{po}$ is the break energy of the observed $\nu F_{\nu}$ gamma-ray spectrum.

The selection was performed by applying the following cuts.

\begin{enumerate}
\item First, we selected GRBs with $T_{90}$, the time for which 90\% of the energy is released, between 2 and 1000 s. This criterion excludes short GRBs ($T_{90}<2s$), and very long GRBs that are superimposed on a varying background and whose $E_{po}$ is difficult to measure accurately.
\item Second, we required reliable spectral parameters. We excluded GRBs with one or more spectral parameters missing. We excluded GRBs with an error on $\alpha$ (the low-energy index of the Band function) larger than 0.5. We excluded a few GRBs with $\alpha <-2.0$ and GRBs with $\beta>\alpha$ because such values suggest a confusion between fitting parameters. We excluded GRBs with large errors on $E_{po}$, which are defined by a ratio of the $90\%$ upper limit to the $90\%$ lower limit larger than 3.5. When the error on $\beta$ in the catalog is lacking or larger than 1.0, we assigned a typical value of –2.3\footnote{This typical value of beta corresponds to the mean values for GRB spectra according to BATSE results, e.g. \cite{Preece2000} and \cite{ Kaneko2006}. It is also very close to the median value beta=-2.26 measured for the {\it Fermi} GRBs , see \cite{Gruber2014}} to $\beta$ and we give no error. In a few cases, the high-energy spectral index in the GRB catalogs ({\it HETE}-2, {\it Fermi}) is incompatible with being $<$-2.0 at the 2 $\sigma$ level, and the catalog gives the energy of a spectral break that is not $E_{po}$. In these cases, we look for $E_{po}$ in the GCN Circulars, and if we cannot find it, then we simply remove the burst from the sample.
\end{enumerate}

Finally, our analysis required GRBs with both well-measured $\gamma$-ray spectral parameters and an optical afterglow light curve. Consequently, we removed GRBs inaccurately localized (typically GRBs not localized by the {\it Swift}-XRT instrument)  because most of the time it prevents ground-based telescope from detecting their optical afterglows.

At the completion of the high-energy selection process, we ended with \nall\ GRBs with a redshift and \ngrbnozall\ GRBs without a redshift.  Then, in a second step, we extracted the short list of GRBs with exploitable optical afterglow light curves.

\subsection{Optical data}
\label{sub_opt}
We collected the afterglow R-band light curves of the pre-selected \nall\ GRBs with a redshift and \ngrbnozall\ GRBs without a redshift. We specifically choose the R-band because it concentrates the largest number of optical measurements. These R-band photometric measurements are issued from published articles and GCN Circulars. References are given in table \ref{tab_refz}.

Then, we used the apparent R magnitude measured 2 hr after the burst uncorrected for extinction as a proxy for the optical flux of the afterglow. The R magnitude is directly interpolated from the available measurements. 
To do so, we required GRBs with good optical follow-up during the first hours after the burst to accurately measure the optical flux of the afterglow.

The choice of the time (2 hr after the trigger) at which we measure the R magnitude results from various constraints.
\begin{enumerate}
\item We want the afterglow to be in its classical slow cooling and decaying regime, yet to be bright enough to permit reliable measurements of the magnitude. 
\item We want to measure the optical flux at a time comparable to the time at which the majority of GRB redshifts are measured (in the first few hours after the trigger), so that the optical flux measurement provides a good indicator of our ability to measure the redshift of the GRBs.
\end{enumerate}

We decided to exclude few GRBs with high visual extinction such that $\rm{A_V^{tot}=A_V^{Gal}+A_V^{Host}>1.2}$, which corresponds to a total extinction in the R-band of about 1 mag. Indeed, GRB afterglows which are strongly absorbed by dust do not provide information on their true optical brightness. This cut is a good trade-off to optimize the number of GRBs for which the afterglow flux is not as polluted by external effects (dust absorption). The galactic extinction $\rm{A_V^{Gal}}$ has been calculated from the dust map of \cite{Schlegel1998} and the host extinctions $\rm{A_V^{Host}}$ are issued from various sources. For some bursts we did not have access to the host extinction. In order to have a rough estimate of the host extinction, we performed a simple linear fit between $\rm{A_V^{Host}}$ and the intrinsic X-ray absorption $\rm{NH_{X,i}}$ derived from the {\it Swift}-XRT catalog. We had 114 GRBs available for this fit. The best-fit gives us the following relation : $\rm{A_V^{Host} = 3.9\times 10^{-23}\times NH_{X,i} + 0.06}$ with a standard deviation of $\sigma\sim~0.34$ mag which we considered to be an acceptable uncertainty in our $\rm{A_V^{Host}}$ estimates.

We also removed those GRBs located at very high redshift since at such redshifts the Ly-$\alpha$ break prevents the observation of the optical afterglow in the R band. The redshift cut was fixed at $z=5.5$ where the Lyman absorption starts to significantly attenuate the R band. 

Finally, for GRB afterglows with only an upper limit of detection, we required that they have R magnitudes that are deeply constrained by large telescopes (at least one 2.0 m telescope). Moreover, these upper limits must be estimated from images taken close to 2 hr after the burst.

\begin{figure}[t!]
\hspace*{-0.4cm}
\includegraphics[trim = 0 150 0 200,clip=true,width=0.54\textwidth]{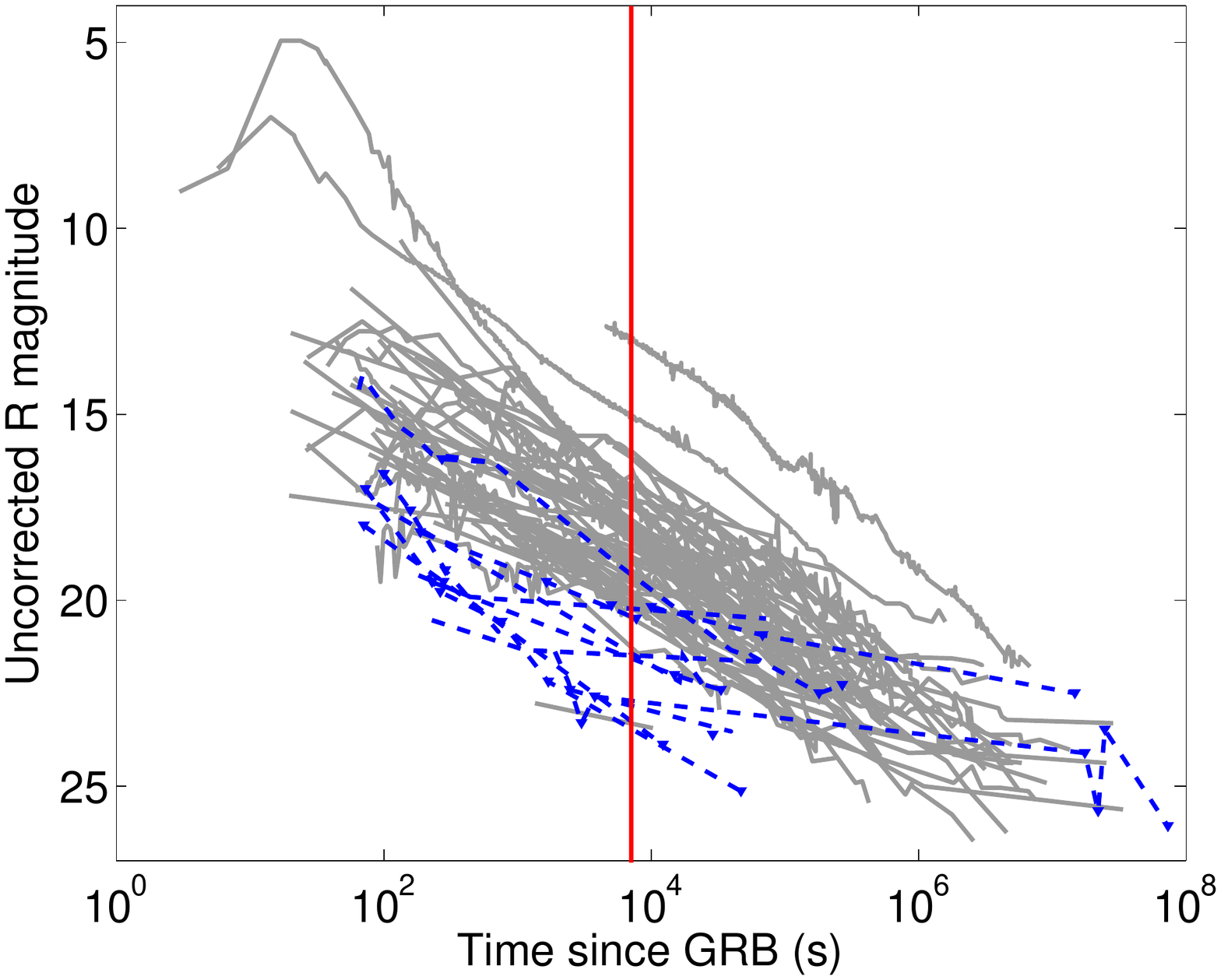}
\caption{R-band optical light curves in the observer frame of the afterglows of \ngrb\ GRBs with a redshift (gray solid line) and \ngrbnoz\ GRBs without a redshift (blue dashed line with triangle for upper limits) considered in this study. The magnitudes are not corrected for galactic and host extinctions. The vertical solid line represents the time (2 hr) at which we estimate the uncorrected R magnitude.
}
\label{fig_afterglow}
\end{figure}

After passing the optical selection criteria, we finally ended with \ngrb\ GRBs with a redshift (75 detections and 1 upper limit) and \ngrbnoz\ GRBs without a redshift (3 detections and 11 upper limits). These 90 GRBs constitute our full sample, which is summarized in table \ref{tab_GRB} for GRBs with a redshift and table \ref{tab_GRB_sansz} for GRBs without a redshift. This sample covers about 15 years of pre-{\it Swift} and {\it Swift} GRB observations (from 1999 to 2014). The afterglow light curves of our complete sample of GRBs can be seen in figure \ref{fig_afterglow}.

\section{Afterglow optical flux and potential biases}
\label{sec_analyse_opt}

\begin{figure}[t!]
\includegraphics[trim = 40 192 0 200,clip=true,width=0.535\textwidth]{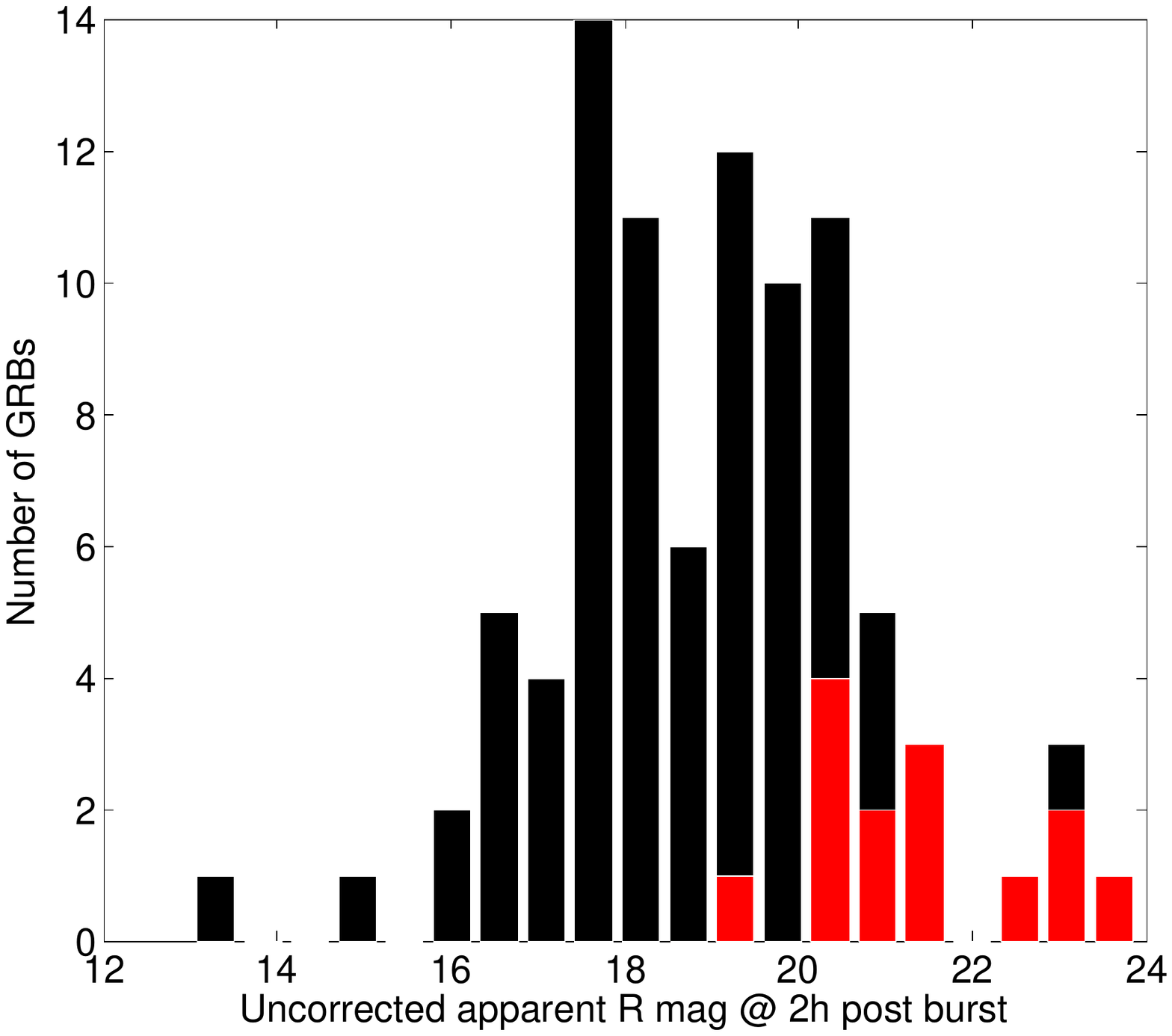}
\caption{Distribution of the afterglow optical flux measured 2 hr after the burst (uncorrected R magnitude). The \ngrb\ GRBs with a redshift are shown in black and the 14 GRBs without a redshift are in red.
}
\label{Rmag_dist}
\end{figure}

The afterglows of our \nallswift\ GRBs span a large range of optical flux from mag R$^{2h}$ = 13.02 to mag R$^{2h}$ = 23.9 (see figure \ref{Rmag_dist}). We noted that GRBs without a redshift have faint optical counterparts, which is not due to high visual extinction since they pass our optical selection criteria. These GRBs may be high-z GRBs or GRBs with sub-luminous afterglows. Our GRBs with a redshift are also distributed over a wide range of redshift from $z=0.168$ to $z=4.11$ and we need to understand whether or not the distribution of the R magnitudes is dominated by the redshift distribution. To verify this hypothesis, we divided our \ngrb\ GRBs with a redshift into three equally populated classes of optical flux. 

\begin{enumerate}
\item The class of {\it bright} GRBs is composed of 26 GRBs with an afterglow R magnitude brighter than $\rm{R=17.9}$.
\item The class of {\it intermediate flux} GRBs is composed of 25 GRBs with an afterglow  R magnitude in the range $\rm{17.9<R\leq19.1}$.
\item The class of {\it faint} GRBs is composed of 25 GRBs with an afterglow  R magnitude weaker than $\rm{R=19.1}$.
\end{enumerate}

We will refer to these classes throughout this paper.

\subsection{Impact of the redshift on the afterglow optical flux distribution}
\label{redshift_dist}

\begin{figure}[h]
\centering
\includegraphics[trim = 40 200 30 190,clip=true,width=0.5\textwidth]{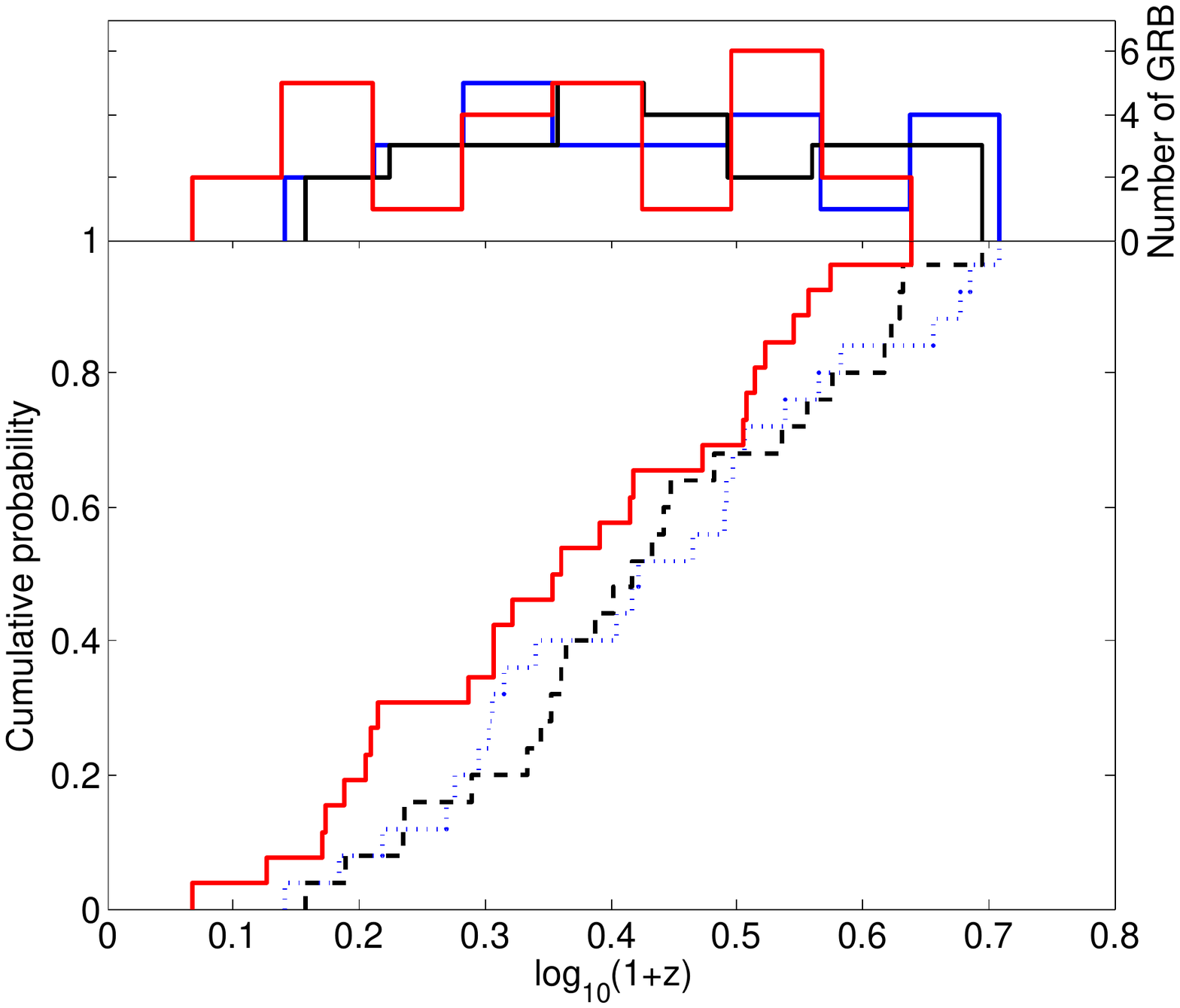}
\caption{Cumulative distribution function (bottom) and histogram (top) of the redshift for the three classes of GRBs. The bright, intermediate, and faint GRB afterglows are indicated by solid red, dashed black, and dotted blue lines, respectively. 
}
\label{z_dist}
\end{figure}

We compared the redshift distribution of our three GRB classes with a Kolomogorov-Smirnov statistical test (KS test; see figure \ref{z_dist}). The KS test clearly reveals that the redshift distributions of the three classes are similar. The results of the statistical tests are summarized in table \ref{tab_stat1}. We conclude that the redshift is not the main driver of our optical flux distribution. This observation is explained in figure \ref{z_vs_LR}, which shows a shift in the GRB population to higher luminosities with the redshift. At large redshifts, the combination of the increased volume and the GRB density evolution allow us to see very luminous GRBs, which are too rare to be visible below $z\sim$1. This luminosity shift nearly compensates for the effect of distance, leading to similar fractions of bright and faint GRBs at all redshifts.

\subsection{Impact of the visual extinction on the afterglow optical flux}
\label{Av_dist}
\begin{figure}[t]
\centering
\includegraphics[trim = 40 190 30 200,clip=true,width=0.50\textwidth]{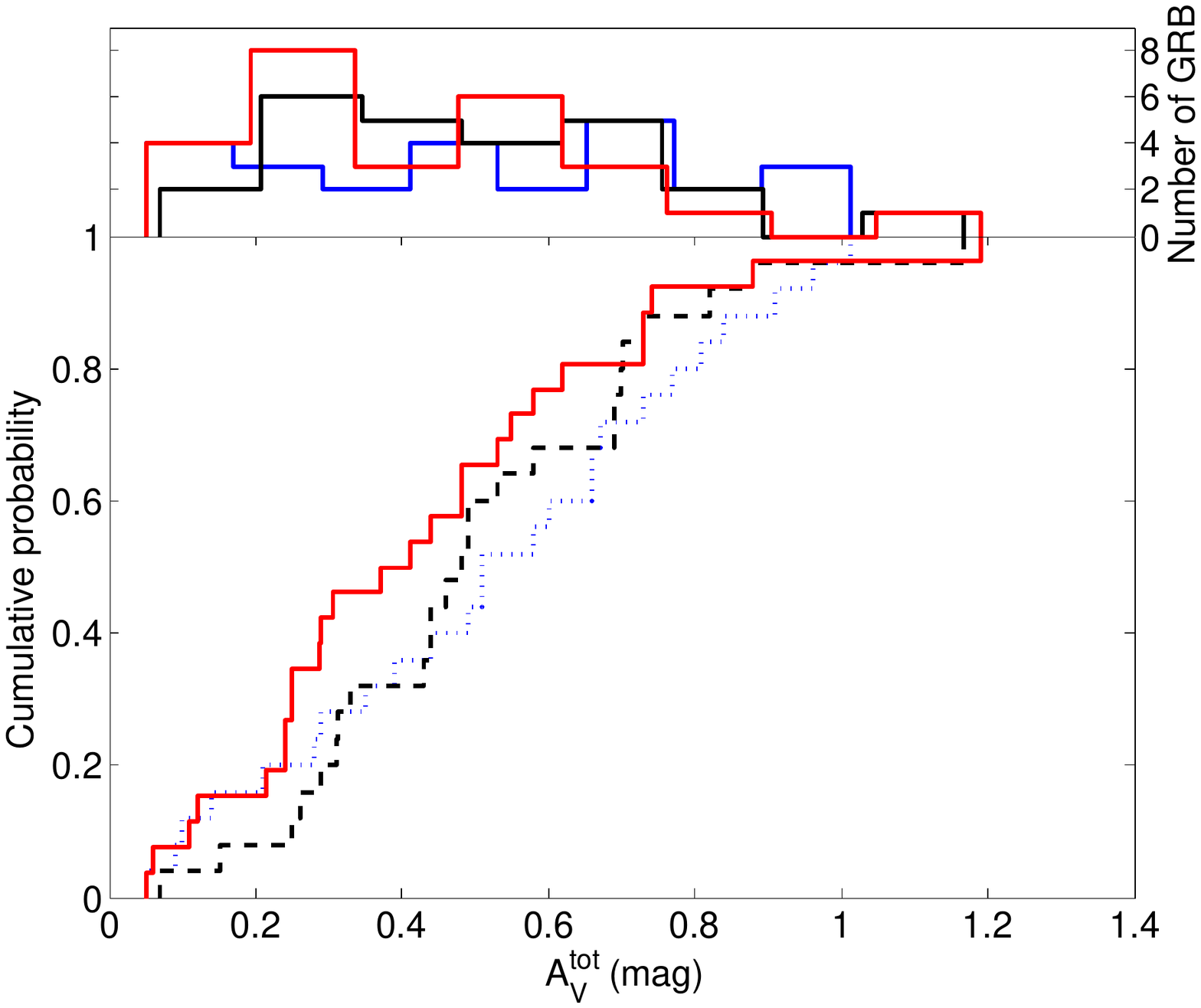}
\caption{Cumulative distribution function (bottom) and histogram (top) of $\rm{A_V^{tot}}$ for the three classes of GRBs. The bright, intermediate, and faint GRB afterglows are indicated by solid red, dashed black, and dotted blue lines, respectively.
}
\label{AV_dist}
\end{figure}

Although we have selected GRBs with relatively low total visual extinction ($\rm{A_V^{tot}<1.2}$), we checked if this parameter could bias our afterglow flux distribution, i.e, whether afterglows with faint optical flux are more obscured by dust. We again performed a KS test to compare the $\rm{A_V^{tot}}$ distributions of our three classes of GRB afterglow flux. We found that the three populations of GRBs are drawn from the same underlying distribution (see figure \ref{AV_dist} and table \ref{tab_stat1}). We conclude that GRBs with faint afterglow optical fluxes are not more obscured than the bright ones, and thus visual extinction does not bias our afterglow optical distribution.

\subsection{Afterglow optical luminosity}
\label{Lopt_dist}
\begin{figure*}[t!]
\begin{minipage}{1.0\textwidth}
\centering
\includegraphics[trim =20 175 70 210,clip=true,width=0.8\textwidth]{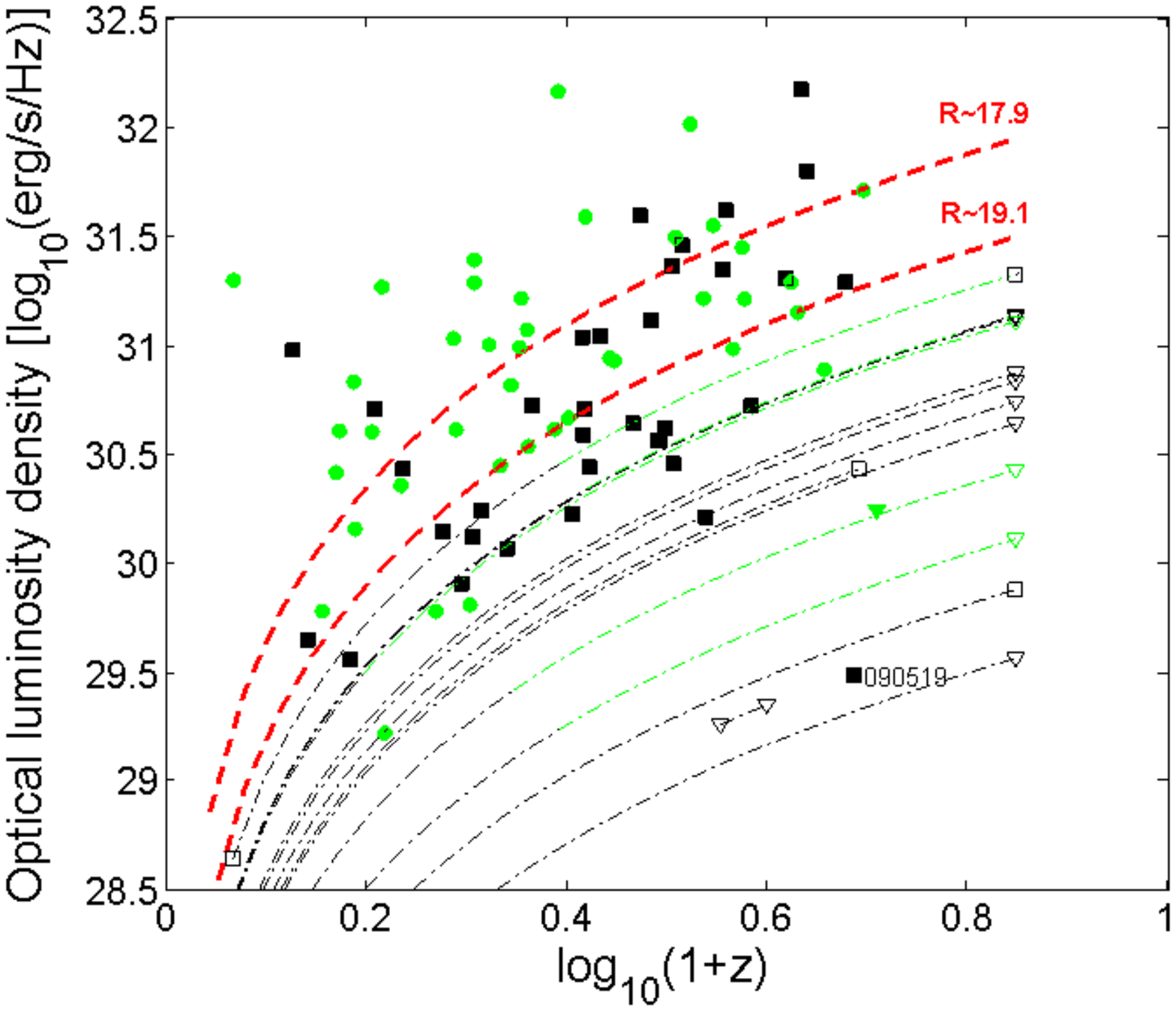}
\end{minipage}
\caption{Intrinsic optical luminosity density as a function of redshift for \ngrb\ GRBs with a redshift and \ngrbnoz\ GRBs without a redshift (dashed-dotted lines). Upper limits are plotted as downward triangles. The red dashed lines indicates the iso-magnitude of the optical afterglow as a function of the redshift (up to $z=6.0$). They define our three classes of optical afterglow flux ({\it bright} are GRBs with R$\le$17.9, {\it intermediate} are GRBs with $17.9 < R\le19.1$, and {\it faint} are GRBs with $R>19.1$). As discussed in section \ref{sec_correlation}, the green circles represent GRBs below the best-fit \eer\ while black squares represent GRBs located above the best-fit \eer. For GRBs without a redshift, the color of the dashed-dotted line indicates whether the GRB is located below the best-fit \eer\ (green) or above it (black).
}
\label{z_vs_LR}
\end{figure*}

As the extrinsic factors (redshift, visual extinction) do not seem to play a major role in the observed optical flux distribution, we investigated the impact of the intrinsic optical luminosity of the afterglow.
We calculated the optical luminosity density (in units of $\rm{erg.s^{-1}.Hz^{-1}}$) two hours after the burst using the following formula:

\begin{equation}
L_R(t_{rest})=\frac{4\pi D_L(z)^2}{(1+z)^{1-\beta_o+\alpha_o}}\times F_{R}(t_{obs})(\frac{\nu_R}{\nu_{obs}})^{-\beta_o}
\end{equation}

where $F_R$ is the optical flux density corrected from the Galactic and host extinction measured at $\rm{t_{obs}=2h}$ after the burst, z is the GRB redshift, $D_L(z)$ is the luminosity distance, $\beta_o$ is the optical spectral index and $\alpha_o$ the optical temporal index, $\nu_R$ is the typical frequency of the R-band (Vega system), and $\nu_{obs}$ is the observed frequency (here $\nu_{obs}=\nu_R$).
For GRBs with no optical detection, we used their optical upper limits to compute $F_R$ and the median values of the $\alpha_o$ ($|\alpha_o^{med}| = 0.975$) and $\beta_o$ ($|\beta_o^{med}| = 0.65$) distributions to estimate an upper limit on their optical luminosity density. 
Finally, for GRBs without a redshift, we also calculated their optical luminosity density as function of the redshift, (considering $0.168<z<6.0$) using the method decribed above. Since their derived  luminosity (or upper limit) depends on the redshift, they produce tracks in the $z-L_R$ plane as shown in figure \ref{z_vs_LR}.

\begin{figure}[t!]
\includegraphics[trim = 40 195 30 200,clip=true,width=0.5\textwidth]{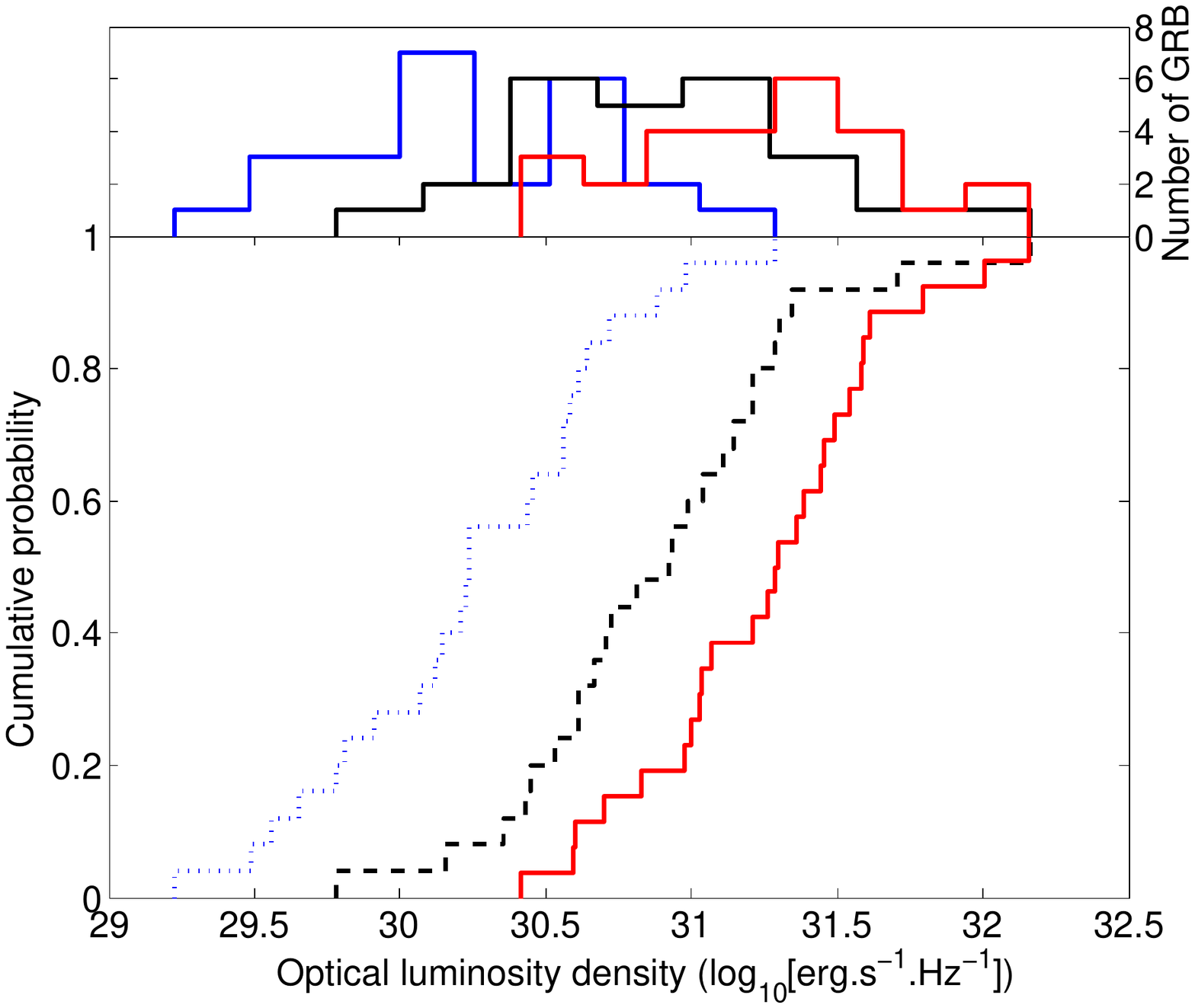}
\caption{Cumulative distribution function (bottom) and histogram (top) of the optical luminosity density (taken 2hr after the burst in the rest-frame) for the three classes of GRBs. The bright, intermediate, and faint GRB afterglows are indicated by solid red, dashed black, and dotted blue lines, respectively.
}
\label{CDF_LR}
\end{figure}

Then, we compared the optical luminosity densities of the three classes of GRB afterglow fluxes ({\it bright}, {\it intermediate}, and {\it faint}; see figure \ref{CDF_LR}). The KS test clearly reveals that GRBs with low optical flux are also less luminous in the optical domain.

We conclude that our afterglow optical flux distribution is strongly shaped by the optical luminosity densities of the GRB afterglows. This result is valid for GRBs with a redshift but, interestingly, GRBs without a redshift also follow this trend. Indeed, they have the afterglows with the lowest optical fluxes (figure \ref{Rmag_dist}), and at any redshift most of them would have had sub-luminous afterglows, see figure \ref{z_vs_LR}. Thus, we conclude that the aferglow optical flux of the GRBs in our sample is very likely dominated by their optical luminosity. As a consequence, a large population of GRBs with sub-luminous afterglows may escape detection in the optical domain, creating a strong bias in the observed afterglow luminosity distribution. In particular, we note that GRBs with $\rm{L_{R}<10^{30}~erg.s^{-1}.Hz^{-1}}$ have no redshift measurement beyond $z\sim1$ (see figure  \ref{z_vs_LR}), except for GRB090519 which is discussed in more details in the Appendix. 

\begin{deluxetable}{cccccccccc}
\tablecolumns{10}
\tablewidth{0pc}
\tablecaption{Results of the different KS tests of section \ref{sec_analyse_opt}.}
\tablecomments{The indicated probabilities correspond to the p-values, i.e, the probability of observing a test statistic as extreme as, or more extreme than, the observed value under the null hypothesis. "F", "I" and "B" refers to the three classes of afterglow optical flux : Faint, Intermediate, and Bright, respectively.  
}
\tablehead{
\colhead{Parameter}    &  \multicolumn{3}{c}{$z$}&  \multicolumn{3}{c}{$A_V$}&  \multicolumn{3}{c}{$L_R$}
}
\startdata
{GRB samples} & F/I & F/B & I/B&  F/I& F/B & I/B & {\bf F/I}& {\bf F/B} &I/B\\[0.5pc]
P-value&  0.30 &   0.47 & 0.88&  0.88 &  0.33 &0.30&  $\mathbf{1.28\times 10^{-3}}$ & $\mathbf{8.06\times 10^{-7}}$ & 0.02 \\
\enddata
\label{tab_stat1}
\end{deluxetable}

\section{Selection effects in the observed GRB population}
\label{sec:sel_effect}

\subsection{The physical picture}
\begin{figure*}[t!]
\centering
\includegraphics[trim = 20 180 60 220,clip=true,width=0.6\textwidth]{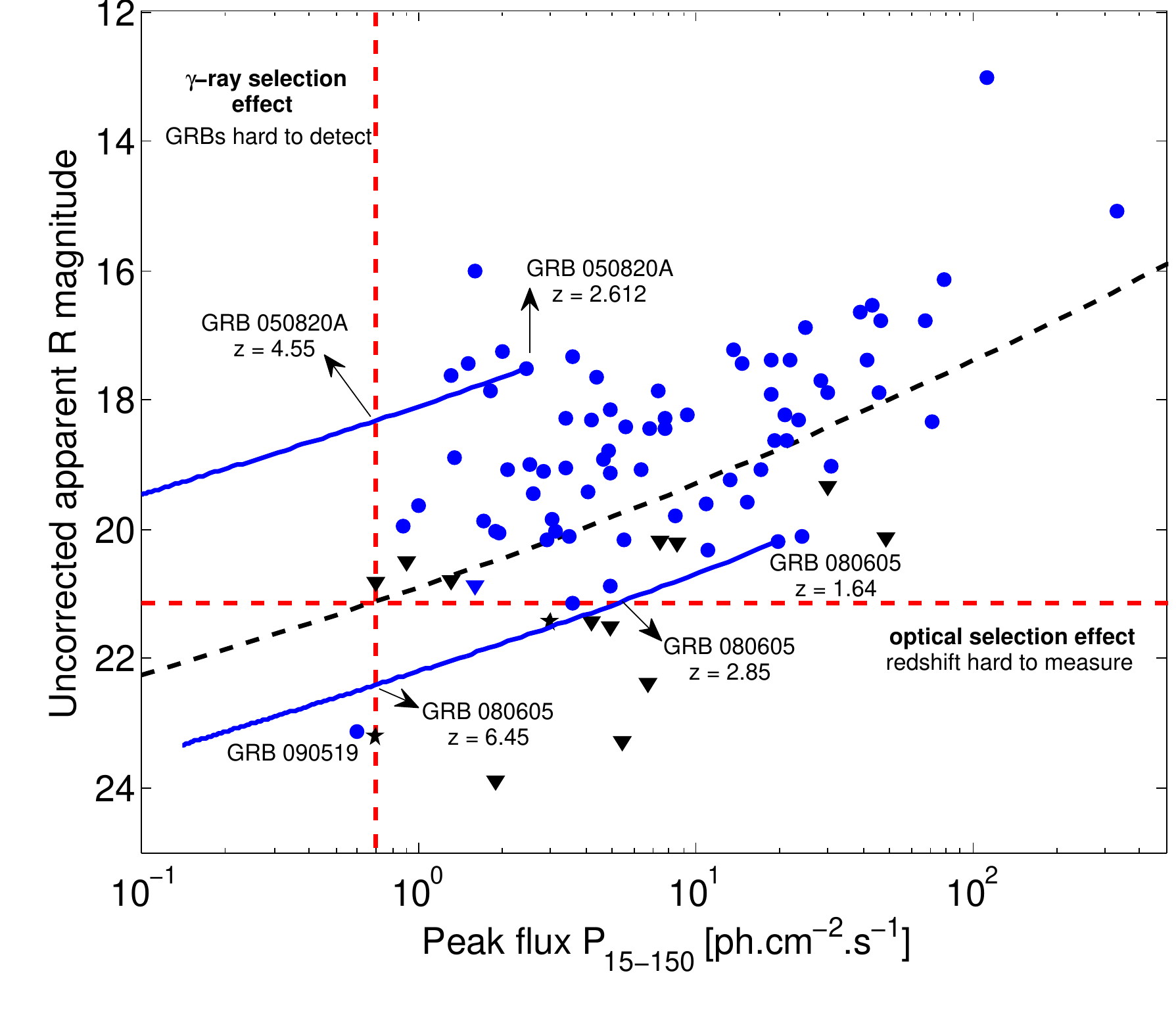}
\caption{Our GRB sample in the Peak flux-R mag plane. GRBs with a redshift measurement are represented in blue (circles for GRBs afterglows with a well-measured R magnitude, downward triangles for optical upper limits). GRBs without a redshift measurement are represented in black (stars for GRBs afterglows with a well-measured R magnitude, downward triangles for optical upper limits). The red dashed lines represent the limiting peak flux and the limiting R magnitude for GRB detection and redshift measurement, respectively. The black dashed line divides the population of GRBs that would undergo the $\gamma$-ray selection effect before the optical one (above this line) from those that firstly suffer from the optical selection effect (below this line).
}
\label{fig:Rmagvspflux}
\end{figure*}

The population of GRBs with a redshift suffers from two different selection effects. The "$\gamma$-ray selection effect" prevents the detection of GRBs with low peak flux while the "optical selection effect" prevents measuring the rest-frame properties of GRBs with faint optical afterglows. In both cases, the final result is that we lose a significant fraction of GRBs for statistical studies dealing with their rest-frame prompt properties. The relative influence of the two selection effects is difficult to assess, but we try to quantify it by comparing the two quantities related to the two selection effects in figure \ref{fig:Rmagvspflux} : 
\begin{enumerate}
\item the observed peak flux (${\rm erg.cm^{-2}.s^{-1}}$) in the {\it Swift} band (15-150 keV) connected to the detectability of a GRB in $\gamma$-rays; and
\item the uncorrected R magnitude of the afterglow related to our ability to measure the redshift of a GRB via the spectroscopy of the optical afterglow.
\end{enumerate}

For each GRB, we simulated the evolution with the redshift of its peak flux and optical afterglow flux. In particular, we checked whether a given GRB located at higher redshift would first disappear from our sample because it becomes undetectable in $\gamma$-rays or because its afterglow is becoming too faint to allow measuring its redshift. Thus, for each GRB, we define $z_\gamma^{max}$ and $z_{opt}^{max}$ as the maximum redshifts that the GRB could have before being limited by the $\gamma$-ray detection threshold ($z_\gamma^{max}$) or the optical threshold related to the redshift measurement ($z_\gamma^{max}$).
To do so, we define a peak flux limit below which the GRB is supposed to be undetectable in $\gamma$-rays at $z = z_{\gamma}^{max}$. We decided to choose the faintest peak flux of our sample (except for GRB090519, which is an exceptional burst), i.e. that of GRB140626A with $P_{15-150{\rm keV}} = 0.7~{\rm ph.cm^{-2}.s^{-1}}$. For the R magnitude limit that defines $z_{opt}^{max}$  we choose the weakest measured R magnitude of our GRB sample with redshift (again we remove GRB090519 from the list), i.e that of GRB090812 with R mag. = 21.15. 

\smallskip

For instance,  GRB080605, with $z_{opt}^{max}\sim2.85$ and $z_\gamma^{max}\sim 6.45$ would be limited by its optical afterglow flux, as shown in figure \ref{fig:Rmagvspflux}. On the contrary, a GRB like GRB 050820A would be limited by the $\gamma$-ray detection with $z_{opt}^{max}>10~$ much\footnote{Note that here we do not take into account the effect of the Lyman break in the R band which appears at $z\sim 5.5-6$. For very high-z GRB ($z>6$), the optical threshold for a redshift measurement has to be preferentially defined based on the infrared afterglow flux (IJHK band).} larger than $z_\gamma^{max}\sim 4.55$ . In the end, we find that about 20$\%$ of our GRBs with redshift would predominantly suffer from optical selection effects if they were located at higher redshift ($z_{opt}^{max}<z_\gamma^{max}$ ), while about 80$\%$ of GRBs would definitely disappear because the GRBs themselves becomes too faint to be detected ($z_{opt}^{max}>z_\gamma^{max}$).

Clearly, the computation of $V/V_{max}$ or other measures of the detectability of GRBs with a redshift have to take into account that some GRBs will be effectively limited by their $\gamma$-ray visibility ($z_\gamma^{max}$), while others ($\sim$ 20 $\%$ of the GRBs with a redshift in our sample) will be firstly limited by their optical afterglow flux  ($z_{opt}^{max}$) and our ability to determine their redshift. In addition, as suggested by figure \ref{Rmag_dist}, we confirm that most of the GRBs in our sample without a redshift are dominated by optical selection effects ($\sim 85\%$ of them) since their peak flux does not prevent them from a $\gamma$-ray detection (see figure \ref{fig:Rmagvspflux}).

We conclude that optical selection effects play a significant role in shaping the observed population of GRBs with a redshift. The next question is as follows : do these optical selection effects create a significant bias in the observed distribution of the rest-frame properties of GRBs?  Below, we discuss the method used to assess the significance of the selection effects associated with the measure of the redshift.

\subsection{A method to assess the significance of optical selection effects}
The impact of measuring the redshift is assessed by comparing the properties of GRBs with different optical fluxes (R magnitudes measured 2 hr after the trigger). 
If GRBs with different optical fluxes show no difference in the distribution of a given parameter, then it is expected that the measure of the redshift, which depends strongly on the optical flux of the afterglow, will not impact the distribution of this parameter. 
If, on the other hand, GRBs with different optical fluxes show significant differences in the distribution of a given parameter, then the measure of the redshift will impact the distribution of this parameter. 

Of course, one limitation of this method is that it cannot be used to measure the biases between the population of GRBs with a redshift and GRBs without one, that is, it can only be used to measure differences between subpopulations of GRBs with a redshift.
The scope of this paper is thus restricted to an evaluation of the prompt GRB properties which may be biased by the measure of the redshift.
A detailed evaluation of the biases between GRBs with and without a redshift would require the construction of a GRB "world model" which takes into account many parameters of the GRBs and their afterglows and their correlations, a task which is beyond the scope of this paper (see, however, \cite{Shahmoradi2013, Kocevski2012}).

\section{Optical selection effects on the rest-frame prompt properties of GRBs}
\label{sec_prompt}

We compare the distributions of various parameters of the prompt emission for the three classes of afterglow flux. The four parameters discussed here are the isotropic $\gamma$-ray energy, $E_{iso}$, the isotropic $\gamma$-ray luminosity, $L_{iso}$, the intrinsic peak energy of the $\nu F_\nu$ $\gamma$-ray spectrum, $E_{pi} = E_{po}\times(1+z)$, and the duration of the burst, $T_{90}^{rest}=T_{90}/(1+z)$. 

\begin{figure*}[t!]
\hspace{-1.9cm}
\begin{minipage}{0.5\linewidth}
\centering
\includegraphics[trim = 0 175 0 200,clip=true,width=1.2\textwidth]{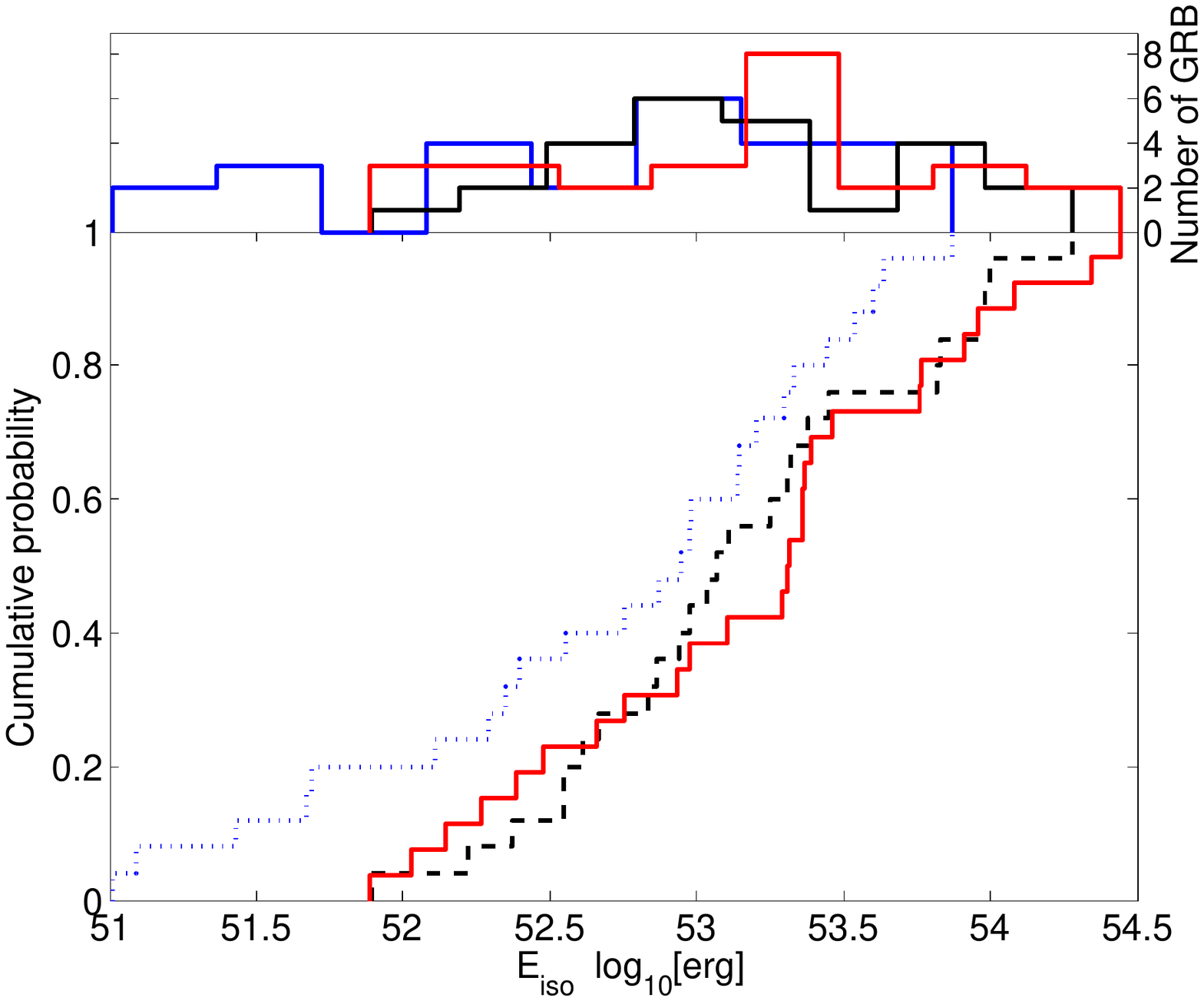}
\end{minipage}
\begin{minipage}{0.5\linewidth}
\centering
\includegraphics[trim = 0 175 0 200,clip=true,width=1.2\textwidth]{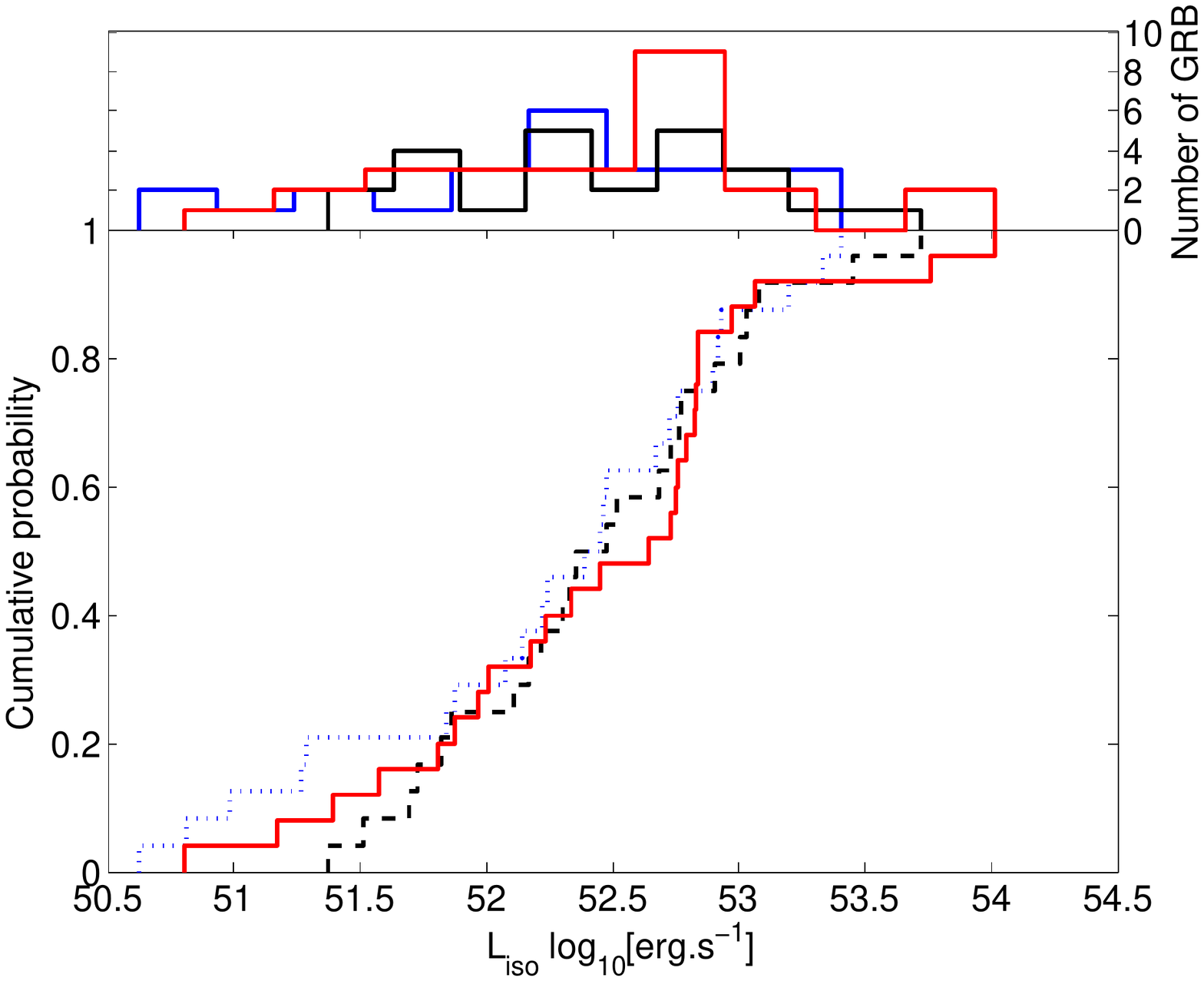}
\end{minipage}
\begin{minipage}{0.5\linewidth}
\hspace{-1.0cm}
\includegraphics[trim = 0 175 0 190,clip=true,width=1.17\textwidth]{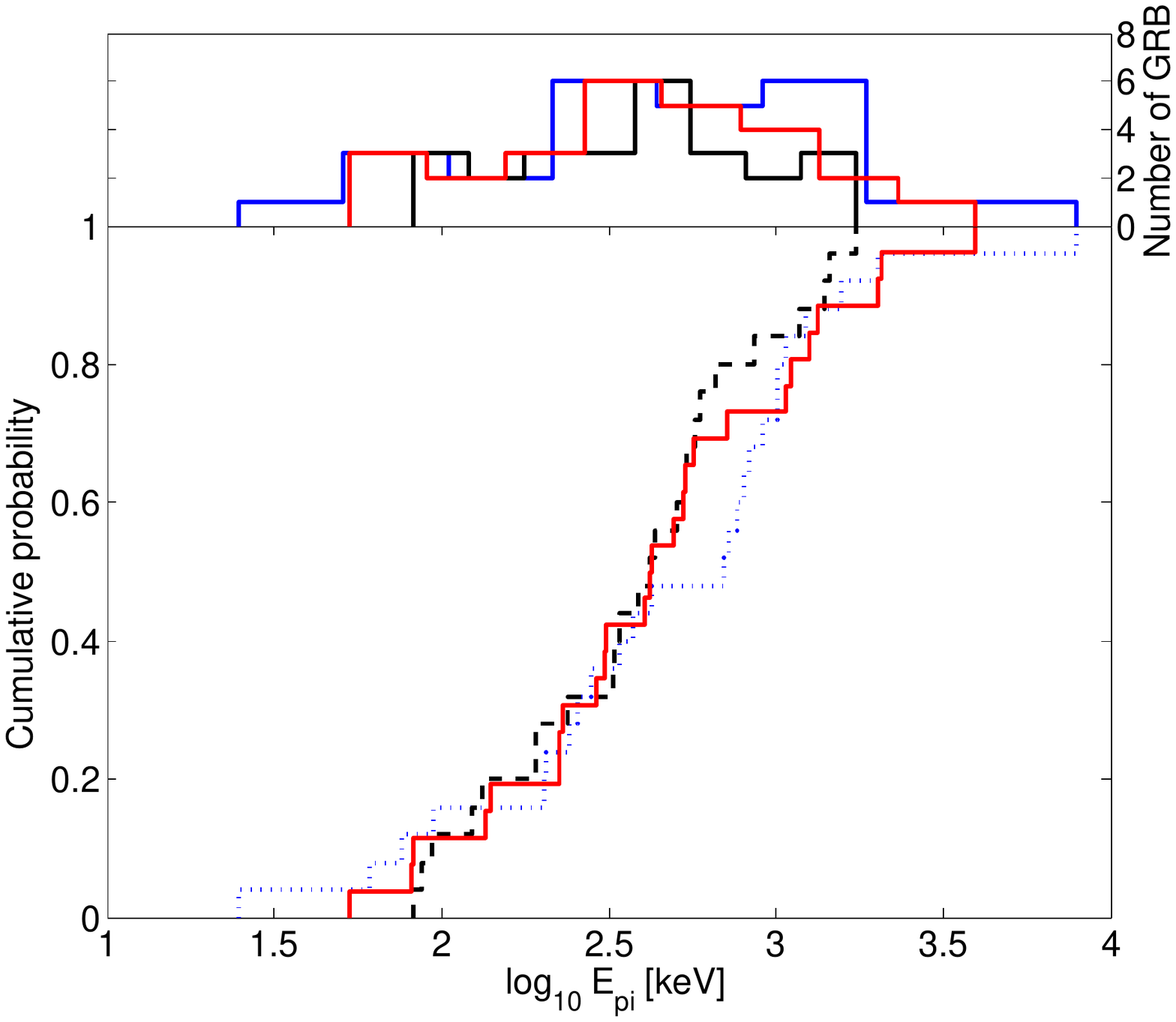}
\end{minipage}
\begin{minipage}{0.5\linewidth}
\includegraphics[trim = 0 175 0 190,clip=true,width=1.17\textwidth]{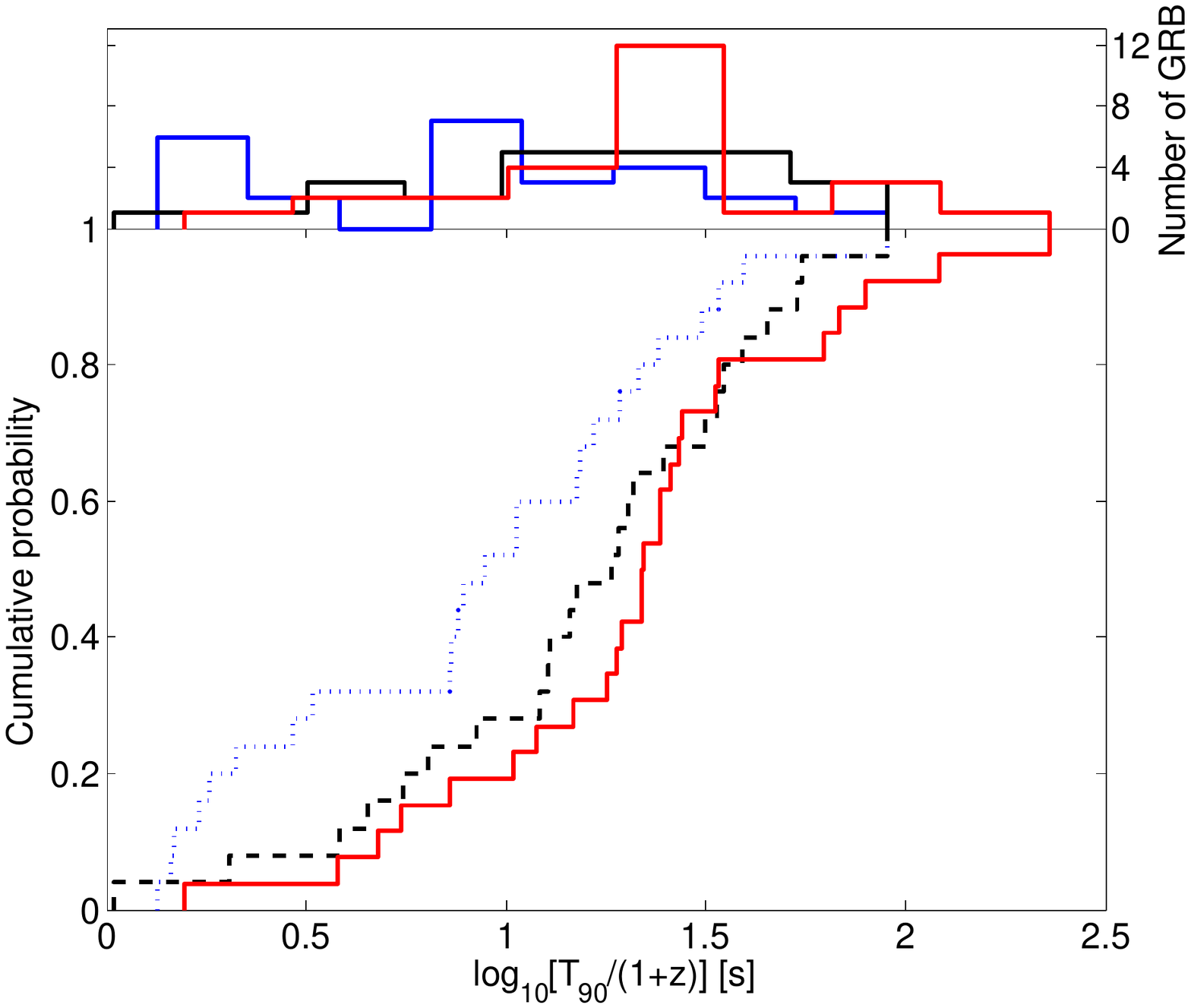}
\end{minipage}
\caption{Top left: cumulative distribution function (bottom) and histograms (top) of the isotropic $\gamma$-ray energy ($E_{iso}$) for the three classes of afterglow optical flux and based on a population of 76 GRBs with a redshift. Top right: same caption for the isotropic $\gamma$-ray luminosity ($L_{iso}$) based on a population of 73 GRBs with a redshift. Bottom left: same caption for the intrinsic peak energy ($E_{pi}$) based on a population of 76 GRBs with a redshift. Bottom right: same caption for the rest-frame burst duration ($T_{90}^{rest}$) based on a population of 76 GRBs with a redshift. All panels: the faint GRBs  are represented with a blue dotted line, the intermediate GRBs with a black dashed line, and the bright GRBs with a red solid line.
}
\label{prompt_par}
\end{figure*}

We performed KS tests to compare the distributions of the four parameters listed above for GRBs in the three classes of afterglow flux (see figure \ref{prompt_par}). These tests are based on the population of 76 GRBs with a redshift when comparing the $E_{iso}$, $E_{pi}$ and $T_{90}^{rest}$ distributions, and only 73 GRBs for the $L_{iso}$ distributions because three GRBs had unsecured peak flux measurements. The results of the KS test are summarized in table \ref{tableprompt}. We found no significant differences between the $L_{iso}$ and $E_{pi}$ distributions. Nevertheless, we noted that the $T_{90}^{rest}$ distributions differ by more than 2.5$\sigma$  between GRBs with faint and bright afterglows. This marginal discrepancy highlights the fact that GRBs with low afterglow fluxes could be shorter on average than GRBs with high afterglow fluxes. Finally, we note that GRBs with faint afterglow fluxes seem to have slightly lower $E_{iso}$ than the rest of the GRB population (intermediate and bright GRB afterglows). This difference is not significant, and so no conclusive statement can be made. The low significance of the KS test applied to $E_{iso}$ can be explained by two reasons. First, the afterglow dynamics are driven by both the kinetic energy of the jet ($E_k = E_{iso}\times\frac{1-\eta}{\eta}$, where $\eta$ is the $\gamma$-ray radiative efficiency) and the density of the shocked ISM. Thus it is not straightforward to connect $E_{iso}$ with the afterglow optical flux. Second to the first order, $E_{iso}$ can be roughly estimated from the averaged $\gamma$-ray luminosity $<L_{iso}>$ integrated over the $T_{90}^{rest}$ duration, $E_{iso}\sim<L_{iso}>\times T_{90}^{rest}$. Since we observed no bias in the $L_{iso}$ distribution but a marginal one in the $T_{90}^{rest}$ distribution, the combination of the two distributions leads to a similar but lower bias in the $E_{iso}$ distribution compared to what we observed for $T_{90}^{rest}$. In the end, no clear trend emerges from these statistical tests and we conclude that the rest-frame prompt properties of GRBs are not significantly biased by optical selection effects

\begin{deluxetable}{ccccccccccccc}
\tablecolumns{13}
\tablewidth{0pc}
\tablecaption{Results of the different KS tests of section \ref{sec_prompt}.}
\tablecomments{The indicated probabilities correspond to the p-values, i.e, the probability of observing a test statistic as extreme as, or more extreme than, the observed value under the null hypothesis. "F", "I" and "B" refer to the three classes of afterglow optical flux, Faint, Intermediate, and Bright, respectively.}
\tablehead{
\colhead {Parameter } &  \multicolumn{3}{c}{$E_{iso}$}&  \multicolumn{3}{c}{$L_{iso}$}&  \multicolumn{3}{c}{$E_{pi}$} &  \multicolumn{3}{c}{$T_{90}^{rest}$}
}
\startdata
{GRB samples}& F/I& F/B &I/B& F/I &  F/B&I/B &  F/I & F/B & I/B &  F/I &  F/B& I/B\\[0.5pc]
P-value&  0.41 & 0.17 & 0.77& 0.62 &  0.73 & 0.99& 0.12&  0.56& 0.99 &  0.12 &  0.02 & 0.53\\
\enddata
\label{tableprompt}
\end{deluxetable}

\section{Optical selection effects on rest-frame prompt correlation}
\label{sec_correlation}
The study of biases in GRB spectral energy correlations has led to many studies concerning $\gamma$-ray selection effects. However, the role of optical selection effects on these correlations has not been explored as much. If such an optical bias exists, then we expect to find a link between the afterglow flux and the positions of GRBs in the corresponding  parameter space of the correlation. We decided to study such a connection using the $\rm{E_{pi}-E_{iso}}$ and $\rm{E_{pi}-L_{iso}}$ relations (respectively reported for the first time by \cite{Amati2002} and \cite{Yonetoku2004}) as they are among the most robust GRB correlations, yet are highly debated in the GRB community.

\subsection{Our GRB sample in the \eep}
\begin{figure*}[t!]
\begin{minipage}{0.5\linewidth}

\includegraphics[trim = 0 175 0 200,clip=true,width=1.0\textwidth]{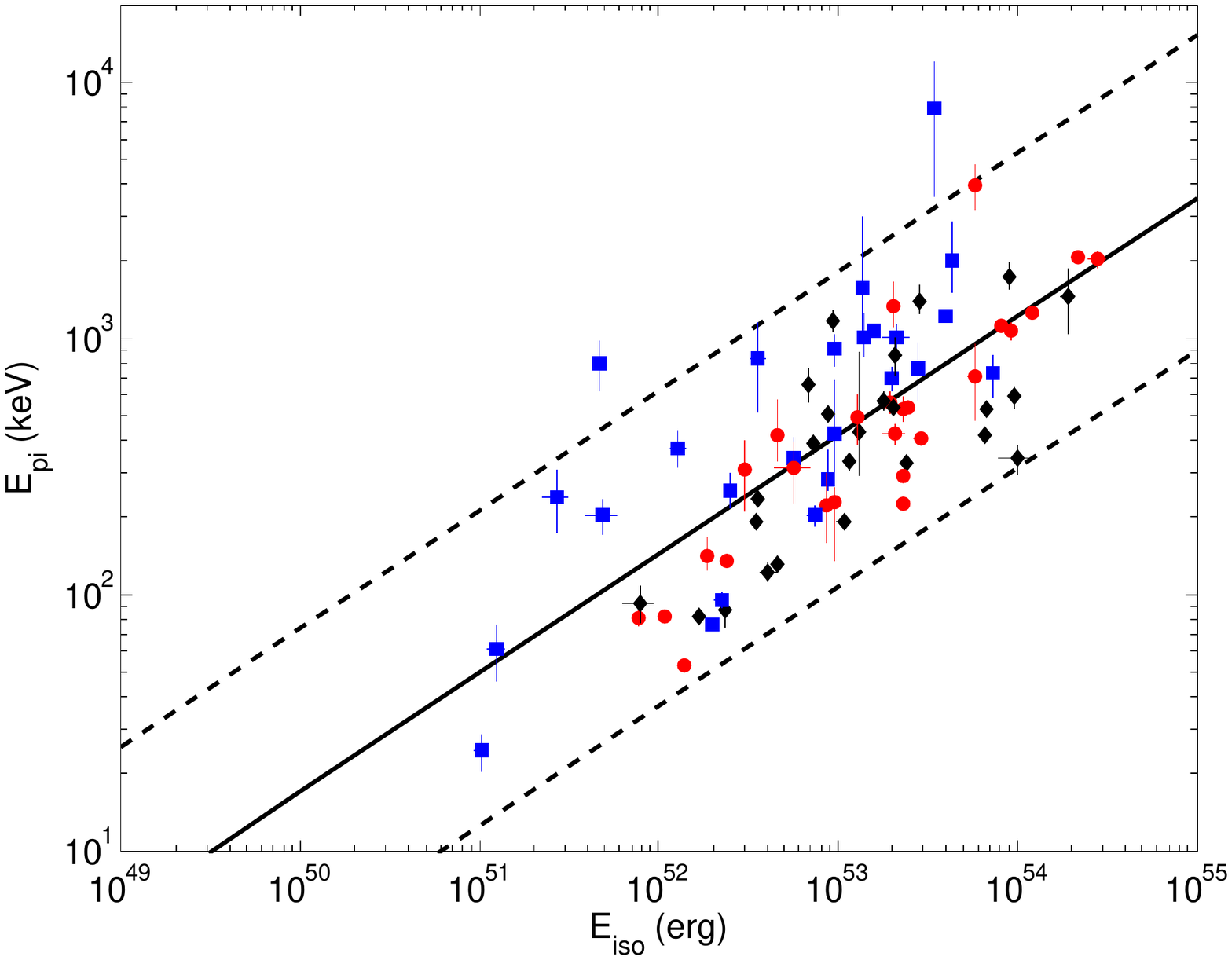}
\end{minipage}
\begin{minipage}{0.5\linewidth}

\includegraphics[trim = 0 175 0 200,clip=true,width=1.0\textwidth]{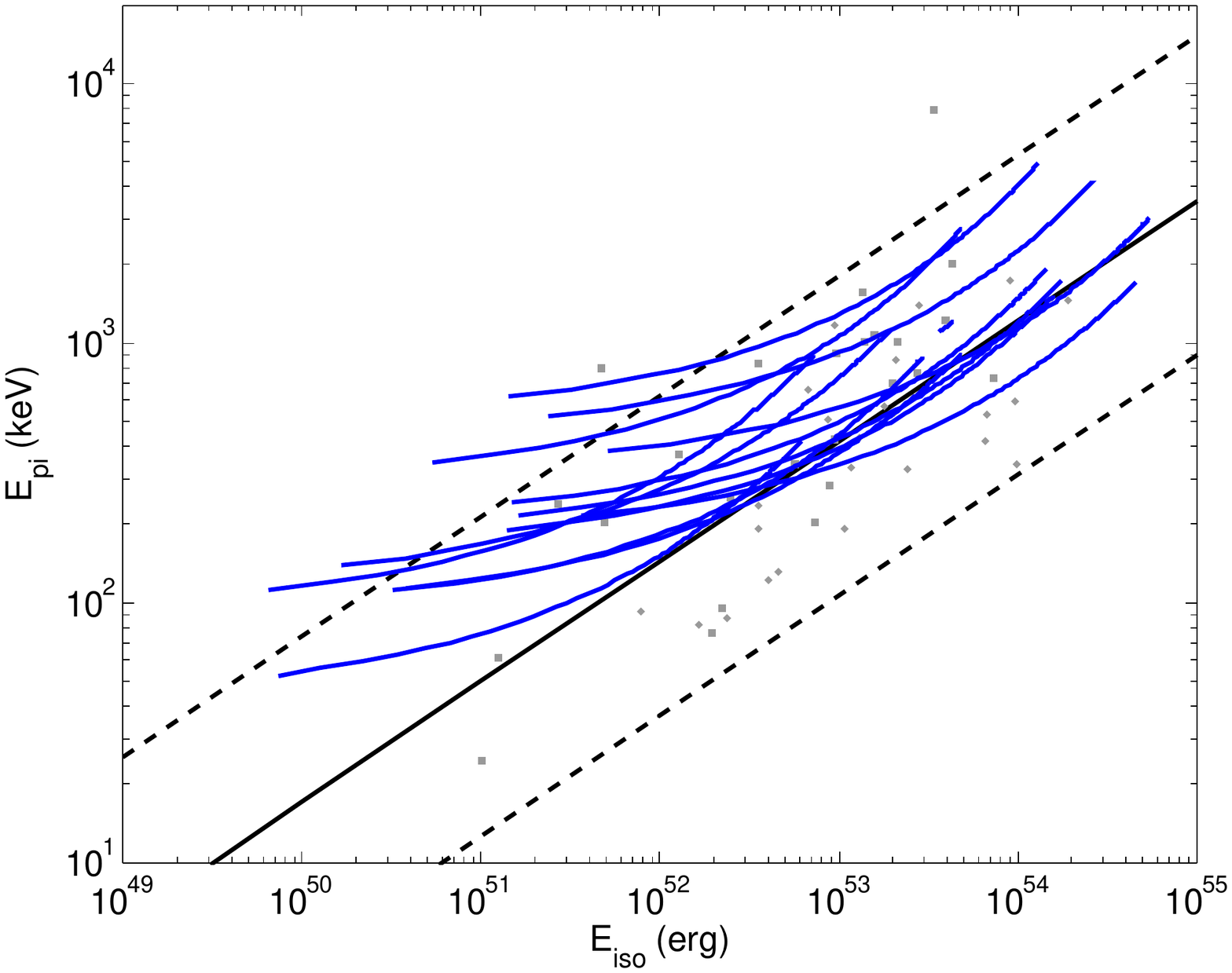}
\end{minipage}
\begin{minipage}{0.5\linewidth}
\hspace*{-0.5cm}
\includegraphics[trim = 0 175 0 200,clip=true,width=1.1\textwidth]{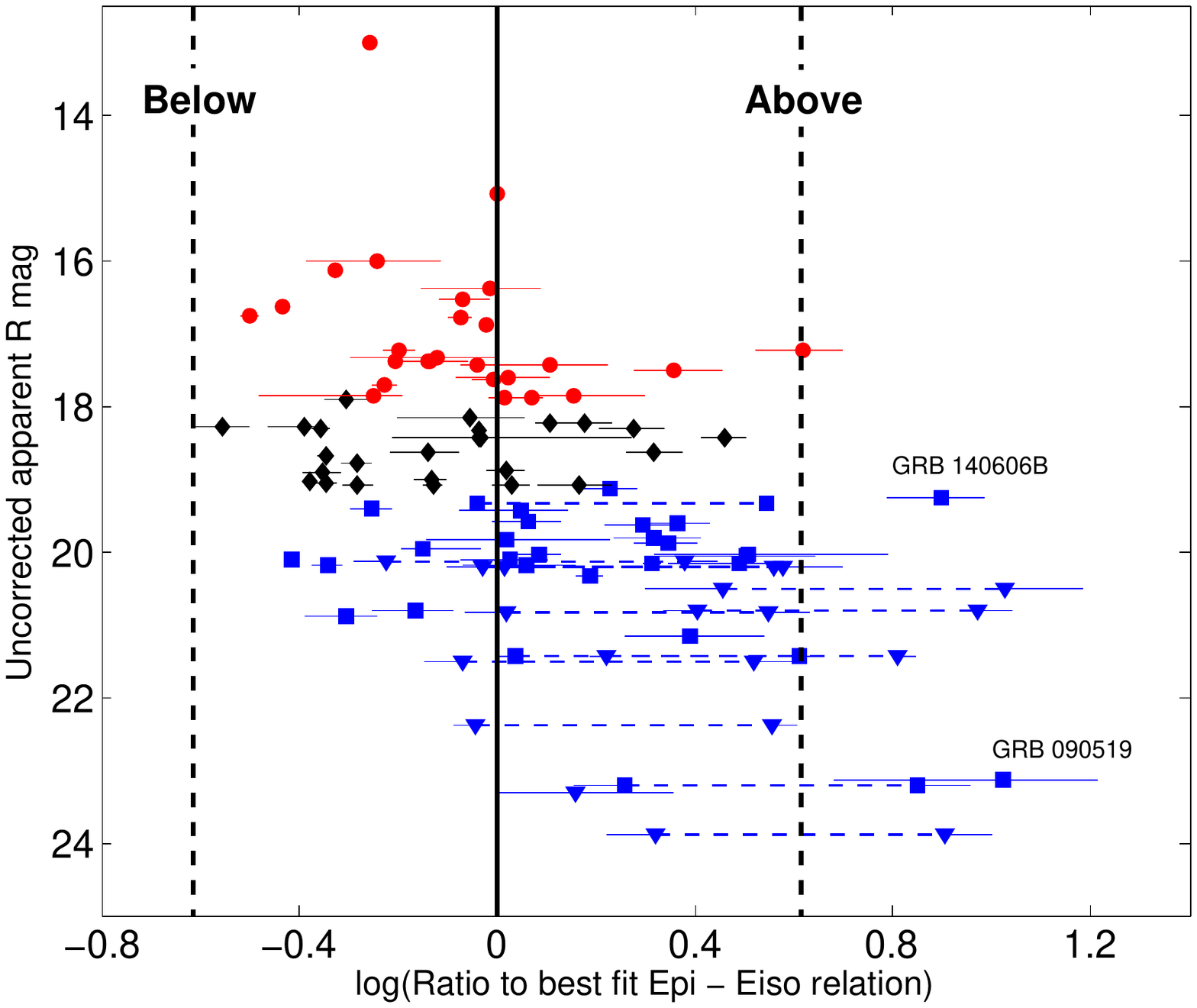}
\end{minipage}
\begin{minipage}{0.5\linewidth}
\hspace*{-0.4cm}
\includegraphics[trim = 0 175 0 190,clip=true,width=1.04\textwidth]{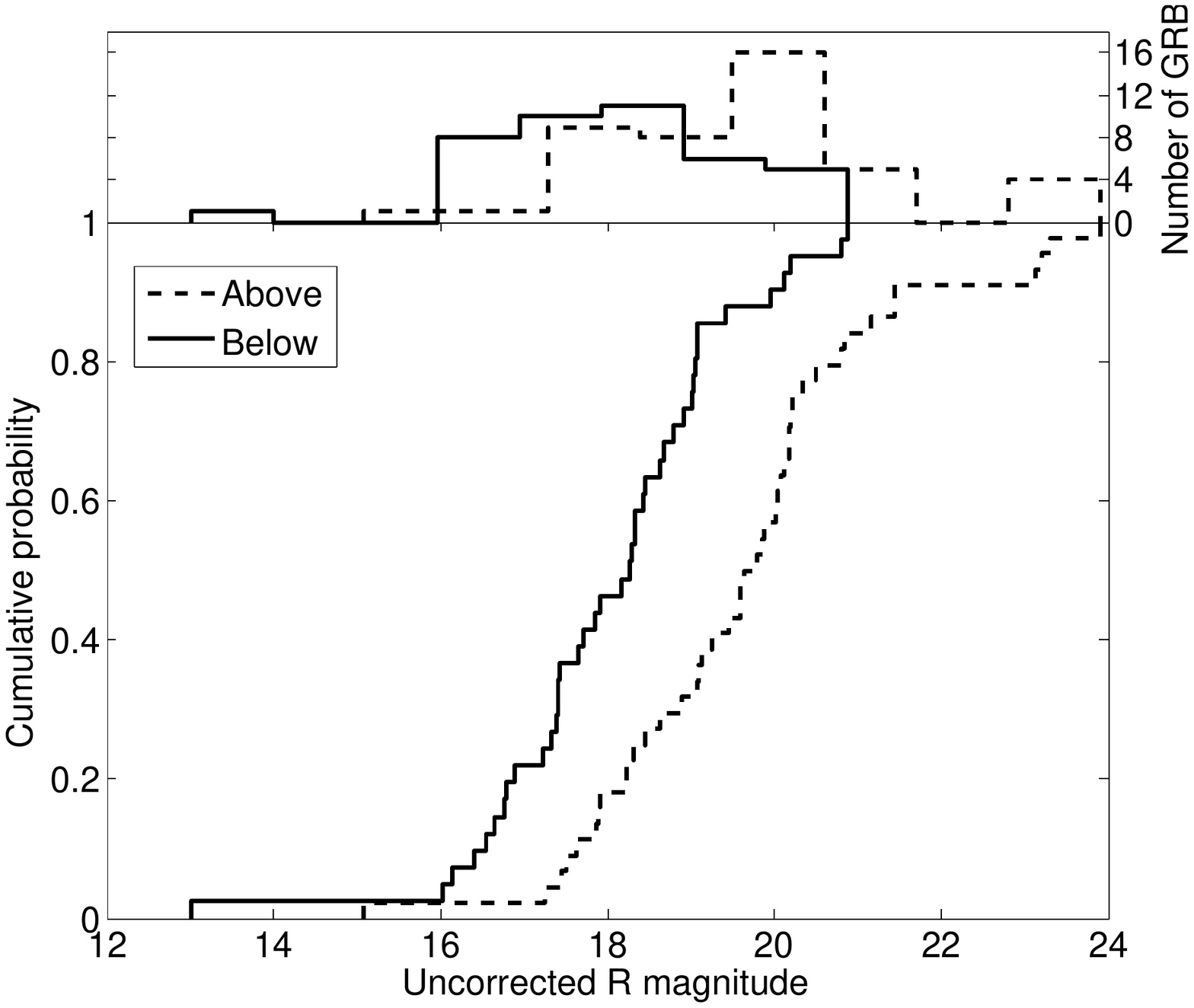}
\end{minipage}
\caption{Top left: distribution of the \ngrb\ GRBs with a redshift in the \eep\ with its 2$\sigma$ limit (dashed line). The red circles corresponds to the population of GRBs with bright afterglows, the black diamonds corresponds to the GRB afterglow with intermediate flux, and the blue squares corresponds to the faint GRB afterglows. Top right: distribution of the \ngrbnoz\ GRBs without a redshift in the \eep\ with its 2$\sigma$ limit (dashed line). They all appear in the class of GRBs with faint afterglows (blue solid line). Bottom left: Distribution of the uncorrected R magnitude as a function of the vertical distance to the best-fit to the \eer\ for our global sample of GRBs (90 GRBs). Bottom right: cumulative distribution function of the uncorrected R magnitude between GRBs located above the best-fit \eer\ (dashed line) and those located below the best-fit \eer\ (solid line). This analysis used a sample of 76 GRBs with a redshift and 8 GRBs without one.
}
\label{fig_amati}
\end{figure*}

As shown in figure \ref{fig_amati}, the \ngrb\ selected GRBs with a redshift follow a standard \eer. 
The best-fit \eer\ for this sample is ${\rm E}_{\rm pi} = 145~{\rm E}_{52}^{0.463}~ {\rm keV}$, where E$_{52}$ is the GRB isotropic energy in units of $10^{52}$ erg.
This best-fit relation is consistent with the \eer\ found by other authors \citep[e.g.][]{Nava2012,Gruber2012}, showing that our sample is not significantly biased for what concerns the distribution of GRBs with a redshift  in the \eep . The dispersion of the points around the best-fit relation along the vertical axis, $\sigma = 0.31$, is also comparable to the values found by \cite{Nava2012} and \cite{Gruber2012} ($\sigma = 0.34$). In the following, we compare the positions of our three classes of GRBs in the \eep\ using the mean distance to best-fit the Amati relation as a criterion for this comparison.

\subsubsection{Comparing the vertical distances with respect to the \eer}
\label{brightness_eep}
The vertical distance is defined as $\rm{log_{10}(E_{pi})} - \rm{log_{10}[best~fit~(E_{pi})]}$, where best-fit ($E_{pi}$) is the value measured on the best-fit relation. In figure \ref {fig_amati}, we show the complete GRB sample (90 GRBs) in a plane displaying the vertical distances to the best-fit \eer\ as a function of the afterglow optical flux. We performed a KS test to compare the vertical distance distributions for our three classes of GRBs with a redshift. The KS tests reveal that GRBs with intermediate and bright afterglow fluxes follow similar distributions, while GRBs with low afterglow fluxes differ by more than $\sim3\sigma$ from GRBs with high afterglow fluxes and by $\sim2.6\sigma$ from GRBs with intermediate afterglow fluxes. Because the faint GRBs seem to behave unlike intermediate and bright GRBs, we compared the vertical distance distributions of these two groups of GRBs (faint versus intermediate+bright). The KS test reveals that the distribution of faint GRBs in the \eep\ is not compatible with that of the intermediate and bright GRBs with a probability of 99.96$\%$ ($\sim 3.7\sigma$). The results of the different statistical tests are summarized in table \ref{tabledist}. This table shows that GRBs with faint afterglows are mostly located in the upper part of the \eep\, compared to the other GRBs. This suggests that an extended population of GRBs with low afterglow fluxes may fill the upper part of the \eep\, even above the 2$\sigma$ limit of the \eer\, but cannot be seen due to optical selection effects that prevent us from measuring their redshift.

\subsubsection{Comparing the afterglow optical flux above and below the best-fit \eer}
In order to confirm the suggestion that GRBs located above the best-fit Amati relation have fainter afterglows than GRBs located below the best-fit Amati relation, we compared the optical flux of the GRBs located above and below the best-fit \eer. In order to include GRBs without a redshift, we calculated their minimum and maximum vertical distances to the best-fit \eer\, which depends on the redshifts (considered here between $0.168<z<6.0$). Then, we only kept those which are strictly located above the best-fit \eer\ (i.e, those for which the minimum and the maximum distances to the best-fit \eer\ are always positive) or strictly below it at any possible redshift. Eight GRBs without a redshift were selected, all of which were located strictly above the best-fit \eer. Therefore, we finally used 84 GRBs in this analysis.

Figure \ref{fig_amati} compares the afterglow optical flux of GRBs located above and below the best-fit \eer. A KS test shows that the two distributions strongly differ with a p-value of $4.77\times 10^{-6}$  ($\sim$ 4.5$\sigma$). The high significance of this test confirms that GRBs located below the best-fit \eer\ have, on average, brighter afterglows than GRBs located above it. The average difference in magnitude between the two groups of GRBs is 1.76. We also noted that this result becomes less significant when we only consider GRBs with a redshift. Thus, GRBs without a redshift seem to confirm the trend that GRBs with low afterglow fluxes are mainly located above the \eer\ and that they can suffer from a lack of redshift measurement.
 
\subsection{Our GRB sample in the \elp}

\begin{figure*}[t!]
\begin{minipage}{0.5\linewidth}

\includegraphics[trim = 0 175 0 200,clip=true,width=1.0\textwidth]{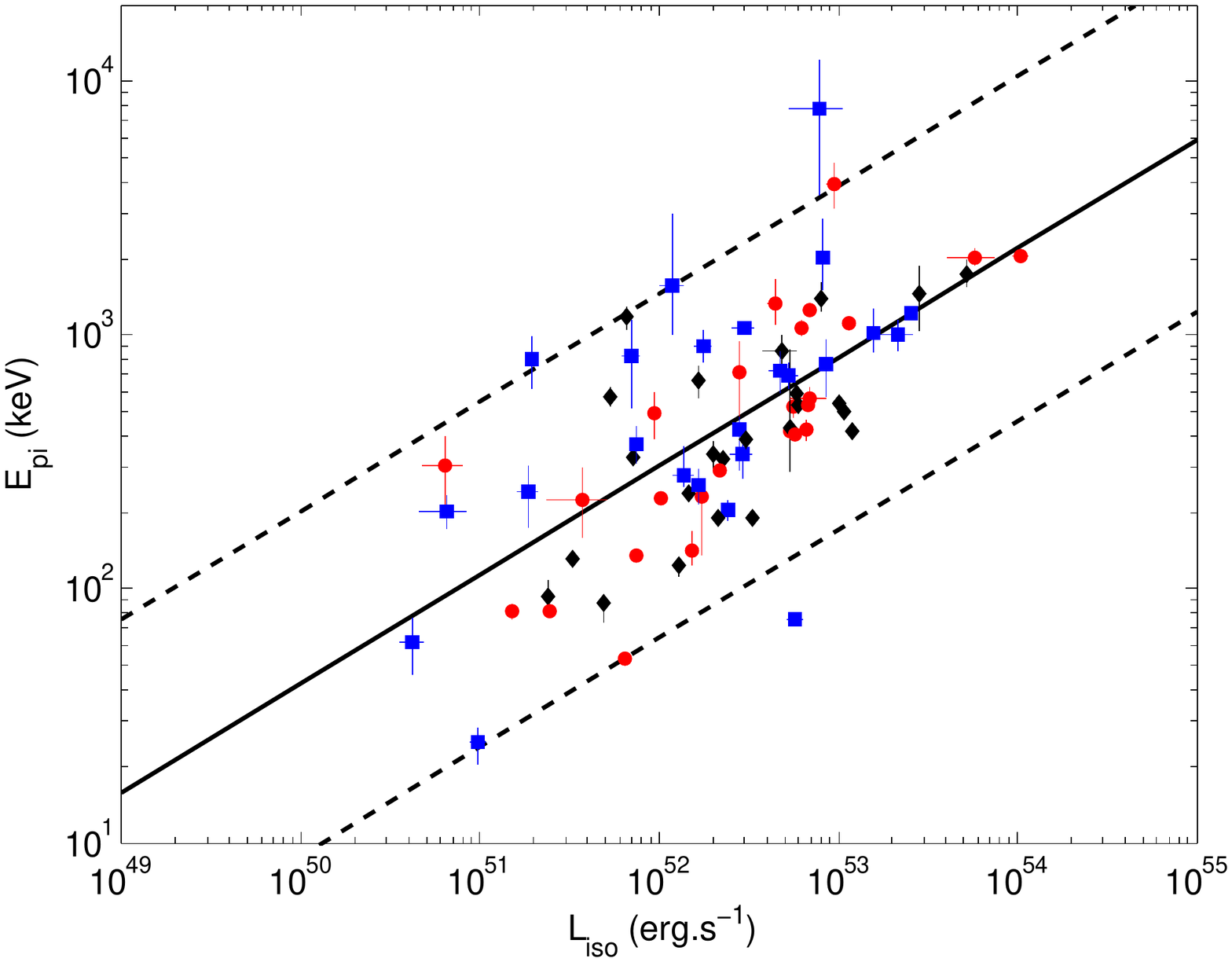}
\end{minipage}
\begin{minipage}{0.5\linewidth}

\includegraphics[trim = 0 175 0 200,clip=true,width=1.0\textwidth]{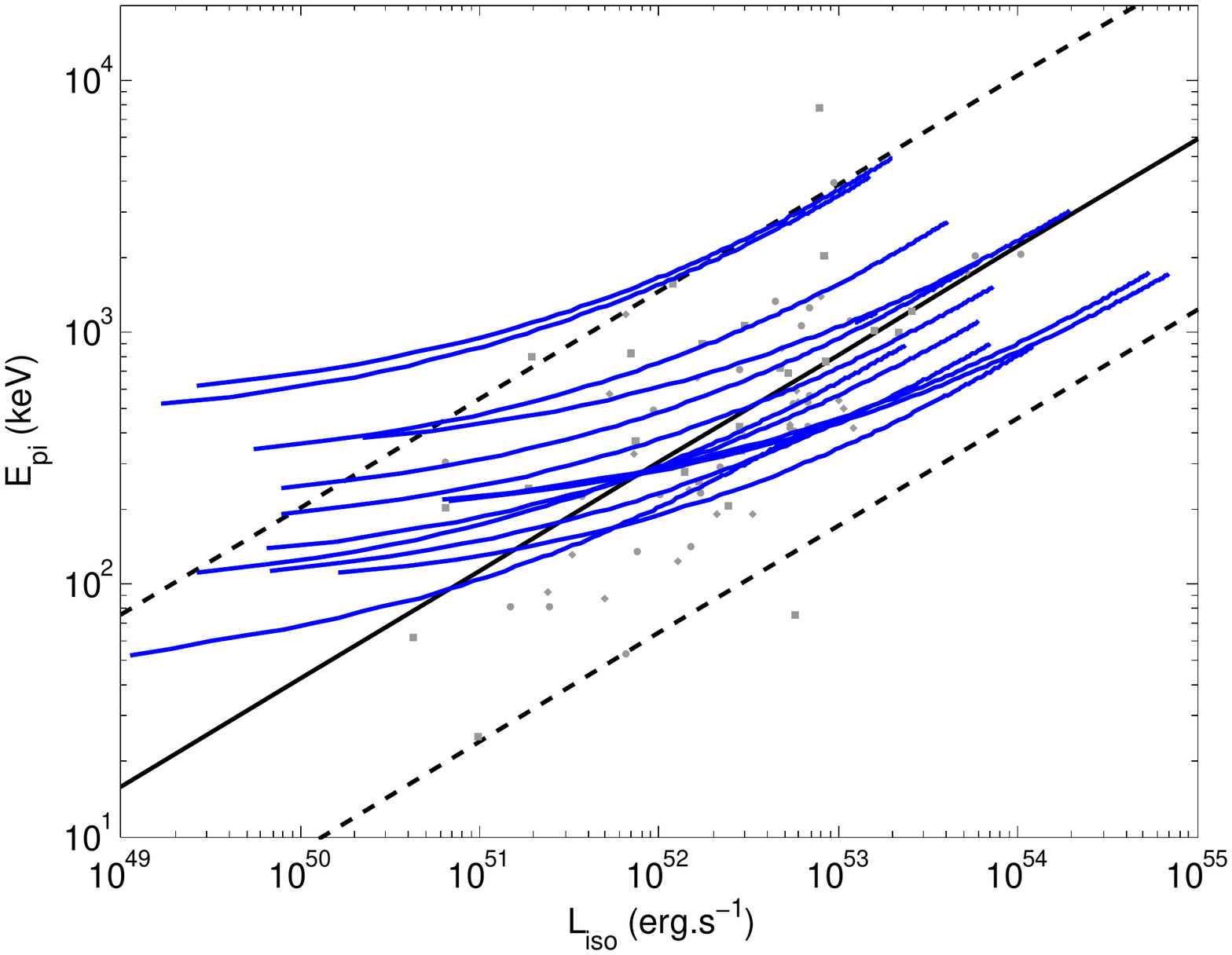}
\end{minipage}

\begin{minipage}{0.5\linewidth}
\hspace*{-0.5cm}
\includegraphics[trim = 0 175 0 200,clip=true,width=1.1\textwidth]{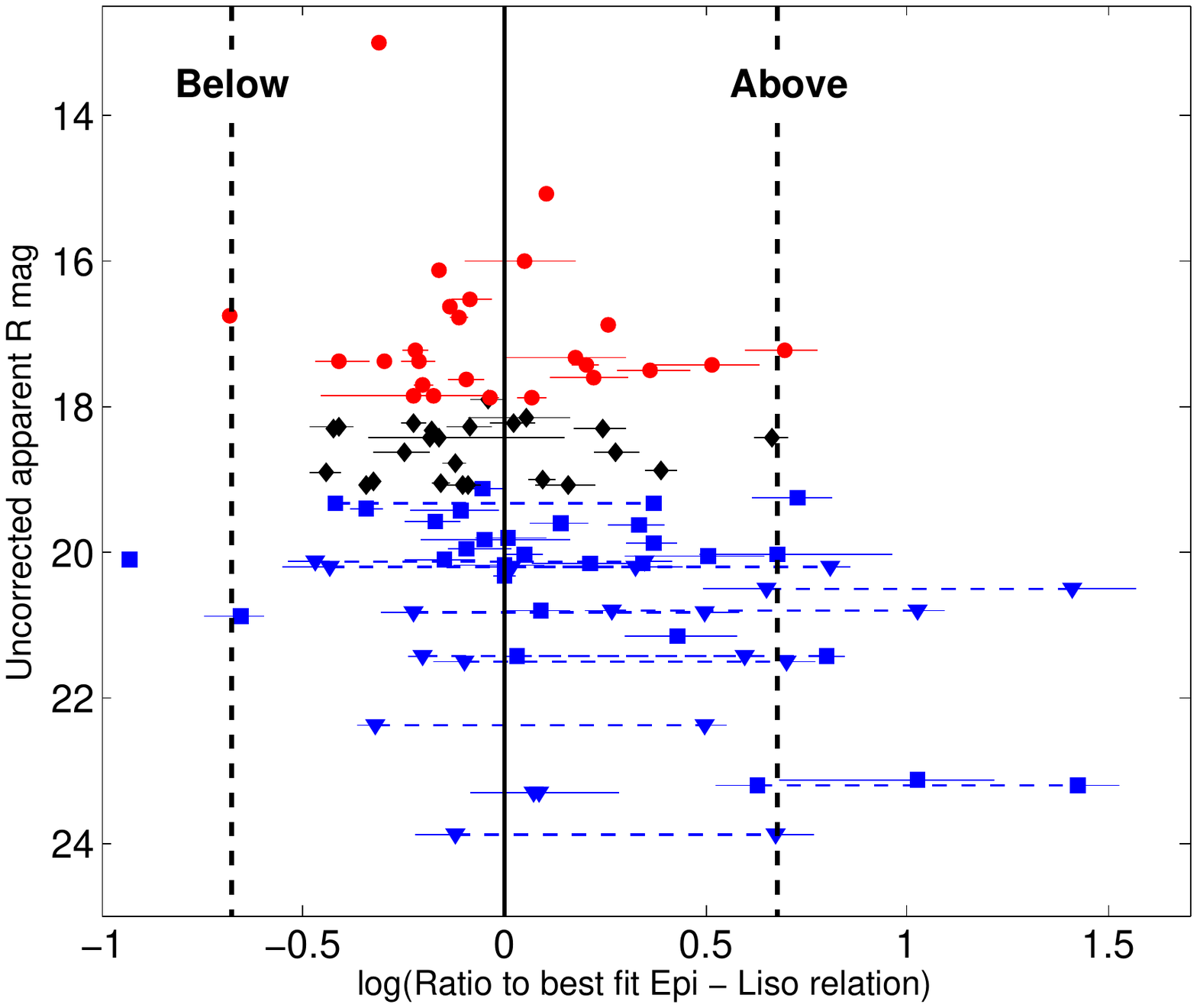}
\end{minipage}
\begin{minipage}{0.5\linewidth}
\hspace*{-0.5cm}
\includegraphics[trim = 0 190 0 190,clip=true,width=1.07\textwidth]{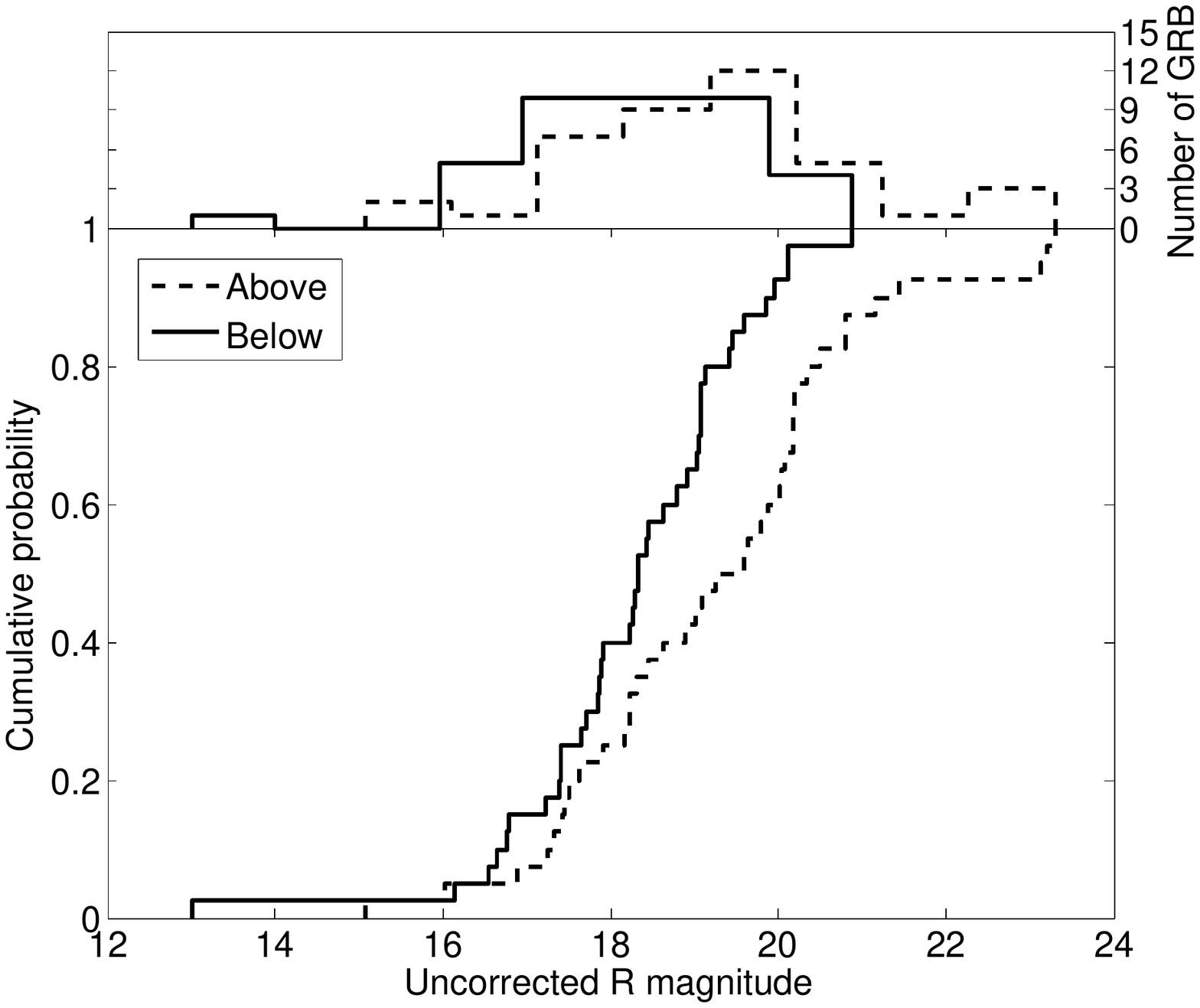}
\end{minipage}
\caption{Same caption as figure \ref{fig_amati}, except E$_{iso}$ changed to L$_{iso}$. Top left: 73 GRBs with a redshift are shown here. Bottom left: the 14 GRBs without a redshift are shown here. Bottom left: our sample of 87 GRBs is shown here. Bottom right: this analysis used a sample of 73 GRBs with a redshift and 6 GRBs without a one.
}
\label{fig_yonetoku} 
\end{figure*}

To compute the standard \elr\ we removed 3 GRBs from our sample of 76 GRBs with a redshift because of unsecured peak flux measurements. The best-fit \elr\ for this sample (73 GRBs) is ${\rm E}_{\rm pi} = 304~{\rm L}_{52}^{0.428}~ {\rm keV}$, where L$_{52}$ is the GRB isotropic luminosity in units of $\rm{10^{52}~erg.s^{-1}}$. 
This best-fit relation is consistent with the \elr\ found by other authors \cite[e.g.][] {Nava2009,Nava2012}. The dispersion of the points around the best-fit relation along the vertical axis, $\sigma = 0.34$, is also comparable with the values found by \cite{Nava2012}, ($\sigma = 0.30$). We then produced the same analysis and the similar figures than for the \eer\ (see figure \ref{fig_yonetoku} and table \ref{tabledist}).

\subsubsection{Comparing the vertical distances to the \elr}

For this analysis, our complete GRB sample is now composed of 87 GRBs since 3 GRBs have unsecured measured peak fluxes. The KS test that compares the vertical distances to the \elr s of our classes of afterglow optical flux reveals that the three populations of GRBs follow nearly the same statistical distribution. Contrary to the \eer\, no clear connection can be determined between the afterglow optical flux and the positions of the GRBs in the \elp.

\subsubsection{Comparing the afterglow optical brightness above and below the best-fit \elr}
In this analysis, we could use 73 GRBs with a redshift and add 6 GRBs without a redshift which are all located above the best-fit \elr, so that we used a total of 79 GRBs. 

The KS test reveals that the two distributions of afterglow optical flux marginally differ with a p-value of $5.0\times 10^{-3}$ ($\sim2.8\sigma$). The significance of this test cannot strictly confirm that GRBs located below the best-fit \elr\ have, on average, brighter afterglows than those of GRBs located above it. The average difference in magnitude between the two groups of GRBs is 0.97.
In conclusion, while we find a clear segregation of GRBs with different optical fluxes in the \eep\, the situation is much less clear in the \elp. The possible origin of this difference is discussed in section \ref{discussion}.
 
\begin{deluxetable}{ccccccc}
\tablecolumns{7}
\tablewidth{0pc}
\tablecaption{Results of the different KS tests of section \ref{sec_correlation}.}
\tablecomments{The indicated probabilities correspond to the p-values, i.e, the probability of observing a test statistic as extreme as, or more extreme than, the observed value under the null hypothesis. "F", "I" and "B" refer to the three classes of afterglow optical flux, Faint, Intermediate, and Bright, respectively.}
\tablehead{
\colhead{Parameter}&  \multicolumn{3}{c}{distance to \eer} &  \multicolumn{3}{c}{distance to \elr}
}
\startdata
{GRB samples}&  F/I & {\bf F/B} & I/B& F/I &  F/B &I/B  \\
 &   &  & & &  & \\[0.5pc]
P-value& $0.01$ & $\mathbf{8.57\times 10^{-4}}$ &0.21& 0.11& 0.28 &0.98 \\
\cline{1-7}\\[-0.80pc]
\cline{1-7}
\\
 Parameter & \multicolumn{6}{c}{Uncorrected R magnitude}\\[0.5pc]
\cline{1-7}\\
 {GRB samples}&  \multicolumn{3}{c}{\bf Above/Below} &\multicolumn{3}{c}{\bf Above/Below}\\
 &  \multicolumn{3}{c}{\bf the \eer} & \multicolumn{3}{c}{\bf the \elr}  \\[0.5pc]
P-value& \multicolumn{3}{c}{$\mathbf{4.77\times 10^{-6}}$} & \multicolumn{3}{c}{$\mathbf{5.0\times 10^{-3}}$}\\
\enddata
\label{tabledist}
\end{deluxetable}


\section{Discussion and conclusion}
\label{discussion}

GRBs with a redshift undergo two types of selection effects: those connected with the detection of the burst in $\gamma$-rays and those connected with the measure of the redshift. 
About 30\% of the GRBs detected with {\it Swift}/BAT have their redshift measured; this small fraction is partly due to a lack of early optical observations (which can nevertheless be compensated for at later times with large observing resources, as in the TOUGH program; \cite{Hjorth2012}) and partly due to the faintness of the optical afterglows, which prevents us from obtaining useful spectra. 
Our study based on a population of 90 GRBs shows that the redshift measure selects the most luminous GRBs. 
When the luminosity dominates over the impact of distance, as is the case for the majority of GRB afterglows, the difficulty to detect faint afterglows biases the GRB optical luminosity function in favor of luminous events. According to our GRB sample, most of the time, the redshifts of GRBs with afterglows less luminous than $\rm{10^{30}~erg.s^{-1}.Hz^{-1}}$ are not measured beyond $z=1$. 

In addition to biasing the distribution of GRB optical luminosities, the measure of the redshift may also impact the observed distribution of prompt GRB parameters if there is a connection between the prompt properties and the optical flux of the afterglow.

\subsection{Redshift and duration}
Our results show that the distributions of \eiso\, \liso\, and \epi\ do not differ much between GRBs with faint and bright optical afterglows. 
Consequently, optical selection effects are not expected to bias these distributions, beyond the biases resulting from the GRB detection in hard X-rays. 

On the contrary, the distribution of $T_{90}^{rest}$ looks different for those GRBs with strong and faint optical afterglows, and thus, this parameter may undergo additional biases due to the measure of redshift. 
According to figure \ref{prompt_par}, the measure of the redshift may select GRBs which are about two times longer than average (i.e. with a larger $T_{90}^{rest}$). This bias is in addition to, and may partially compensate for, the biases resulting from the $\gamma$-ray selection effects discussed by \cite{Kocevski2013}. A detailed analysis of the $\gamma$-ray light curves would be helpful to assess more precisely the differences between the light curve of GRBs with bright and faint optical afterglows.

\subsection{Redshift and GRB correlations}
We have found a significant correlation between the optical magnitudes of GRBs and their locations in the \eep. 
In our sample, GRBs with a large optical flux concentrate in the region located below the best-fit \eer, while GRBs with low optical flux preferentially fill in the region located above the best-fit \eer , like a majority of GRBs without a redshift, which have very faint optical afterglows. 
This optical bias could explain the apparent contradictory results obtained between GRBs with a redshift, which seem to follow the \eer\ relation quite well, and the whole GRB population, which seems to contain a significant fraction of outliers. 
Our observations also show that GRBs with a redshift underestimate the true width of the \eer\ correlation.
We propose an explanation for this observation in the next section.

On the contrary, we  observe no clear correlation between the afterglow optical brightness and the positions of GRBs in the \elp. 
This means that the \elr\ does not suffer as much from optical selection effects.
If biases affect the observed \elr\, then they should mostly come from the $\gamma$-ray detections. 

\subsubsection{Why Do GRBs below and above the Best-fit \eer\ have different R magnitudes?}
Figure \ref{z_vs_LR} clearly shows that the excess of bright optical afterglows among GRBs located below the best-fit \eer\ is due to a bunch of GRBs with $\rm R < 17.9$ and $\rm z < 1.5$ ($\rm log(1+z) < 0.4$). 
Above $z=1.5$, on the other hand, the bright GRBs are equally distributed with respect to the \eer . 
This led us to study the \eer\ below and above redshift $z=1.5$. 
We find that the best-fit moves from ${\rm E}_{\rm pi} = 128~{\rm E}_{52}^{0.397}~ {\rm keV}$ for GRBs with  $\rm z < 1.5$ to ${\rm E}_{\rm pi} = 211~{\rm E}_{52}^{0.400}~ {\rm keV}$ for GRBs with  $\rm z \geq 1.5$.
Assuming that GRBs below and above $z=1.5$ follow different \eer s,  the bright GRBs are now distributed equally above and below the best-fit \eer\ for GRBs with $\rm z < 1.5$. Using the same method as in section \ref{sec_correlation}, we compared the R magnitudes of GRBs with a redshift located below their best-fit \eer\ to those located above it and now found an insignificant result with a KS test p-value = 0.0329. Thus, we observe that the correlation between optical brightness and the location of GRBs in the \eep\ disappears. 

We conclude that there is no real difference between GRBs located below and above the \eer. Instead, the \eer\ evolves with redshift or changes with GRB luminosity, leading to the observed correlation between the location of GRBs in the \eep\ and the magnitude of their afterglows when a single \eer\ is considered in the full redshift range.
A similar conclusion was proposed by \cite{Li2007} and  \cite{Lin2015}.

Based on the present statistics, it is not possible to determine whether the change of the \eer\ below and above  $\rm z = 1.5$ is due to an evolution of the \eer\ with the redshift or to different relations for GRBs with small and large \eiso . Considering that the location of GRBs in the \eep\ is an intrinsic property, like their energy \eiso\, we tend to favor the second possibility.

\subsubsection{Using the \eer\ for cosmological purposes}

The simplest way to use the \eer\ for GRB standardization is to consider that it as an intrinsic GRB property that does not evolve with redshift or with the properties of the GRBs. 
However, this simple view is being challenged by growing evidence that the \eer\ is a boundary in the \eep\ and not a true correlation (e.g. \cite{Heussaff2013}), and by the possibility that the \eer\ evolves with redshift (\cite{Li2007,Lin2015}; see, however, \cite{Ghirlanda2008}).

We have shown here that the selection effects due to the measure of the redshift cannot be neglected when discussing the \eer\ as a genuine rest-frame prompt property of long GRBs or when attempting to use it for cosmology. 
In addition, our study suggests that GRBs which can be calibrated with Type Ia supernobvae (at redshifts $z < 1.5$) are not representative of the GRB population at higher redshift, which raises important concerns for GRB cosmology based on the \eer (see for example \cite{Liang2008}).

We conclude that GRB standardization is a complex issue that cannot rely on the construction of a GRB Hubble diagram simply based on the measure of \epi\, and the redshift and an "ideal" \eer . 
It requires a better understanding of the nature of the \eer\ (boundary or true correlation), of its dependence on GRB parameters, of its possible evolution with redshift, and of the biases resulting from current observations.

\subsection{Conclusion}

Large samples of GRBs with a redshift are required in order to obtain a correct understanding of the intrinsic properties of GRBs and their afterglows. 
Such samples suffer from combined selection effects due to the need to detect the GRBs and the need to measure their redshift.
The connections that exist between the prompt emission and the afterglow imply that these selection effects will impact both the observed properties of the prompt emission and those of the afterglow.

Here, we have studied the potential impact of measuring the redshift on the observed properties of the prompt emission.
According to our results, the redshift measurement slightly alters the observed distribution of GRB durations, leading to a longer-than-average selection of GRBs. 
A stronger effect is observed for the \eer . 
We have found a correlation between the location of a GRB in the \eep\ and the optical R magnitude of its afterglow, we interpret this observation as being due to the dependence of the \eer\ on the GRB energy or its evolution with redshift.



\begin{deluxetable}{ccccccccccc}
\tablecolumns{10}
\tablewidth{0pc}
\tablecaption{Parameters of the GRB Sample with a redshift}
\tablecomments{(1) GRB name; (2) redshift; (3) rest-frame peak energy of the gamma-ray spectral energy distribution; (4) isotropic gamma-ray energy; (5) isotropic gamma-ray luminosity; (6) duration, in seconds, during which 90$\%$ of the burst fluence was accumulated starting by the time at which 5$\%$ of the total fluence has been detected; (7) $R_{mag}$ is the apparent R magnitude 2 hr after the burst not corrected from the galactic and host extinctions ($\blacktriangledown$ indicates it is an upper limit on R mag); (8) $L_R^{rest}$ is the optical luminosity density taken 2 hr after the burst (in the rest-frame); (9) galactic extinction from \cite{Schlegel1998}; (10) host extinction (*) indicates that the host extinction is estimated from the $\rm{NH_{X,i}}$ measurement), otherwise the references for the host extinction are listed in table \ref{tab_refz}; (11) the references are listed in table \ref{tab_refz}.
}
\tablehead{
\colhead{GRB }& \colhead{ z} & \colhead{ $E_{pi}$}&\colhead{ $E_{iso}$}&\colhead{ $L_{iso}$}&\colhead{ $T_{90}$} & \colhead{ $R_{mag}$}&\colhead{ $log_{10}(L_R^{rest})$} &\colhead{ $\rm{A_V^{gal}}$}& \colhead{ $\rm{A_V^{host}}$}& {Ref. z}\\
\colhead{ } & \colhead{ }& \colhead{(keV)}&\colhead{ ($10^{52} erg$) }& \colhead{ ($10^{52} erg.s^{-1}$)}& \colhead{ (s)}&\colhead{ (mag)}& \colhead{($\rm{erg.s^{-1}.Hz^{-1}})$}&\colhead{ (mag)}& \colhead{ (mag)}& \colhead{ }\\
\colhead{ (1)}& \colhead{ (2)} & \colhead{ (3)} & \colhead{ (4)}& \colhead{ (5)} &\colhead{ (6)}&\colhead{ (7) }& \colhead{ (8)}& \colhead{ (9)}&\colhead{ (10)} &\colhead{ (11)}}
\startdata
 990123 &  1.60 &  $2030.0^{+161.0}_{-161.0}$ & $278.0^{+31.5}_{-31.5}$ & $57.56^{+16.9}_{-16.9}$ & 63.3 & $17.90$ & 31.03&0.05&$\sim 0$$~^{a)}$& A.\\[0.5pc]
 990510&  1.619&  $423.0^{+42.0}_{-42.0}$ & $20.6^{+2.9}_{-2.9}$ & $6.68^{+0.56}_{-0.56}$ & 67.6 & $17.40$ & 31.58&0.66&0.22$~^{a)}$& B.\\[0.5pc]
 990712&  0.434&  $93.0^{+15.0}_{-15.0}$ & $0.78^{+0.15}_{-0.15}$ & $0.24^{+0.02}_{-0.02}$ & 30 & $18.64$ & 29.78&0.11&0.15$~^{b)}$& C.\\[0.5pc]
 020124&  3.198&  $339.6^{+44.0}_{-44.0}$ & $99.8^{+21.0}_{-21.0}$ & $2.01$ & 51.2 & $18.29$ & 31.29&0.16&0.28$~^{a)}$& D.\\[0.5pc]
 020813&  1.25&  $592.9^{+60.2}_{-60.2}$ & $95.9^{+3.6}_{-3.6}$ & $5.79$ & 87.9 & $17.91$ & 30.99&0.36&0.12$~^{a)}$& E.\\[0.5pc]
 021004&  2.33&  $310.5^{+84.0}_{-84.0}$ & $5.68^{+1.26}_{-1.26}$ & \nodata & 48.9 & $16.40$ & 32.01&0.20&0.26$~^{a)}$& F.\\[0.5pc]
 021211&  1.01&  $94.8^{+6.8}_{-6.8}$ & $2.23^{+0.23}_{-0.23}$ & \nodata & 4.2 & $20.19$ & 29.81&0.09&$\sim 0$$~^{a)}$& G.\\[0.5pc]
 030328&  1.52&  $327.3^{+22.6}_{-22.6}$ & $24.0^{+0.73}_{-0.73}$ &$2.28$  & 138.3 & $18.79$ & 30.66&0.15&$\sim 0$$~^{a)}$& H.\\[0.5pc]
 030329&  0.168&  $82.2^{+1.5}_{-1.5}$ & $1.07^{+0.02}_{-0.02}$ &$0.25$& 25.9 & $13.02$ & 31.29&0.08&0.54$~^{a)}$& I.\\[0.5pc]
 040924&  0.859&  $75.9^{+2.5}_{-2.5}$ & $1.96^{+0.14}_{-0.14}$ & $5.71$ & 3.4 & $20.12$ & 29.78&0.19&0.16$~^{a)}$& J.\\[0.5pc]
 041006&  0.716&  $82.2^{+3.1}_{-3.1}$ & $1.66^{+0.05}_{-0.05}$ & \nodata & 22.1 & $18.68$ & 30.35&0.71&0.11$~^{a)}$& K.\\[0.5pc]
 050416A&  0.6535&  $24.8^{+3.8}_{-4.5}$ & $0.10^{+0.01}_{-0.01}$ & $0.09^{+0.01}_{-0.01}$ & 2.4 & $20.87$ & 29.22&0.10&0.19$~^{a)}$& L.\\[0.5pc]
 050525A&  0.606&  $135.1^{+2.7}_{-2.7}$ & $2.41^{+0.04}_{-0.04}$ & $0.75^{+0.42}_{-0.42}$ & 8.8 & $17.40$ &  30.60&0.32&0.26$~^{c)}$& M.\\[0.5pc]
 050820A&  2.612&  $1326.7^{+343.4}_{-224.1}$ & $20.3^{+1.43}_{-1.43}$ & $4.42^{+0.17}_{-0.17}$ & 26 & $17.51$ & 31.61&0.15&0.065$~^{d)}$& N.\\[0.5pc]
 050922C&  2.198&  $417.5^{+162.8}_{-85.7}$ & $4.56^{+0.15}_{-0.15}$ & $5.35^{+0.24}_{-0.24}$ & 5 & $17.87$ & 31.36&0.34&0.07$~^{e)}$& O.\\[0.5pc] 
 060115&  3.53&  $280.9^{+86.1}_{-27.2}$ & $8.87^{+0.78}_{-0.78}$ & $1.38^{+0.19}_{-0.19}$ & 139.6 & $19.97$ & 30.88&0.44&$\sim 0^*$& P.\\[0.5pc]
 060908&  1.1884&  $425.3^{+264.1}_{-130.3}$ & $9.48^{+0.38}_{-0.38}$ & $2.81^{+0.23}_{-0.23}$ & 19.3 & $19.85$ & 30.08&0.12&0.09$~^{a)}$& Q.\\[0.5pc] 
 061007&  1.261&  $1065.4^{+81.4}_{-81.4}$ & $91.41^{+1.16}_{-1.16}$ & $6.23^{+0.16}_{-0.16}$ & 75.3 & $17.44$ & 31.21&0.07&0.66$~^{f)}$& R.\\[0.5pc] 
 061121&  1.314&  $1402.3^{+208.3}_{-166.6}$ & $28.21^{+0.41}_{-0.41}$ & $8.04^{+0.18}_{-0.18}$ & 81.3 & $18.63$ & 30.72&0.15&0.28$~^{a)}$& S.\\[0.5pc]
 070612A&  0.617&  $305.6^{+95.4}_{-95.4}$ & $3.0^{+0.17}_{-0.17}$ & $0.06^{+0.02}_{-0.02}$ & 368.8 & $17.45$ & 30.67&0.17&0.46$^*$& T.\\[0.5pc] 
 071010B&  0.947&  $87.6^{+7.8}_{-1.4}$ & $2.35^{+0.05}_{-0.05}$ & $0.49^{+0.02}_{-0.02}$ & 35.7 & $18.28$ & 30.61 &0.03&$0.18^*$& U.\\[0.5pc]
 071020&  2.145&  $1014.3^{+252.0}_{-167.0}$ & $14.06^{+0.61}_{-0.61}$ & $15.81^{+0.56}_{-0.56}$ & 4.2 & $19.81$ & 30.61&0.21&0.28$~^{a)}$& V.\\[0.5pc] 
 080319B&  0.937&  $1261.0^{+27.1}_{-25.2}$ & $120.36^{+1.49}_{-1.49}$ & $6.90^{+0.14}_{-0.14}$ & 66 & $16.89$ & 31.03&0.04&0.07$~^{a)}$& W.\\[0.5pc] 
 080411&  1.03&  $525.8^{+71.1}_{-54.8}$ & $23.24^{+0.09}_{-0.09}$ & $5.59^{+0.12}_{-0.12}$ & 56 & $16.54$ & 31.36&0.11&$0.28^*$& X.\\[0.5pc]
 080413A&  2.433&  $432.6^{+449.7}_{-144.2}$ & $12.97^{+0.37}_{-0.37}$ & $5.38^{+0.19}_{-0.19}$& 19 & $18.43$ & 31.19&0.51&$\sim 0$$~^{d)}$& Y.\\[0.5pc] 
 080413B&  1.10&  $140.7^{+27.3}_{-16.8}$ & $1.85^{+0.06}_{-0.06}$ & $1.51^{+0.06}_{-0.06}$ & 8 & $17.39$ & 31.00&0.12&$\sim 0$$~^{a)}$& Z.\\[0.5pc] 
 080605&  1.64&  $768.2^{+198.0}_{-198.0}$ & $27.60^{+0.41}_{-0.41}$ & $8.53^{+0.26}_{-0.26}$ & 20 & $20.20$ & 30.40&0.44&0.47$~^{g)}$& AA.\\[0.5pc]
 080721&  2.591&  $1747.0^{+241.3}_{-212.5}$ & $91.00^{+7.58}_{-7.58}$ & $52.31^{+4.51}_{-4.51}$ & 16.2 & $18.24$ & 31.34&0.34&0.35$~^{d)}$& AB.\\[0.5pc] 
 080804&  2.2045&  $697.2^{+78.4}_{-78.4}$ & $19.80^{+1.01}_{-1.01}$ & $5.28^{+0.68}_{-0.68}$ & 34 & $20.03$ & 30.43&0.05&0.06$~^{g)}$& AC.\\[0.5pc] 
 080810&  3.355&  $3957.3^{+801.7}_{-801.7}$ & $58.58^{+2.55}_{-2.55}$ & $9.42^{+0.94}_{-0.94}$ & 106 & $17.25$ & 31.79&0.09&0.16$~^{d)}$& AD.\\[0.5pc] 
 081007&  0.5295&  $61.2^{+15.3}_{-15.3}$ & $0.12^{+0.01}_{-0.01}$ & $0.04^{+0.01}_{-0.01}$ & 12 & $19.45$ & 29.56&0.05&$\sim 0$$~^{a)}$& AE.\\[0.5pc] 
 081008&  1.9685&  $493.3^{+107.7}_{-107.7}$ & $12.70^{+0.59}_{-0.59}$ & $0.93^{+0.07}_{-0.07}$ & 185.5 & $17.62$ & 31.59&0.31&0.46$~^{a)}$& AF.\\[0.5pc] 
 081121&  2.512&  $564.6^{+58.1}_{-58.1}$ & $19.57^{+1.43}_{-1.43}$ & $6.92^{+1.57}_{-1.57}$ & 42 & $17.65$ & 31.52&0.17&$0.13^*$& AG.\\[0.5pc] 
 081222&  2.77&  $538.1^{+36.1}_{-36.1}$ & $20.28^{+0.42}_{-0.42}$ & $10.13^{+0.26}_{-0.26}$ & 24 & $18.45$ & 31.21&0.07&$\sim 0$$~^{a)}$& AH.\\[0.5pc]
 090102&  1.543&  $1073.8^{+45.9}_{-45.9}$ & $15.96^{+0.70}_{-0.70}$ & $2.98^{+0.43}_{-0.43}$ & 27 & $20.17$ & 30.22&0.15&0.45$~^{a)}$& AI.\\[0.5pc] 
 090418A&  1.608&  $1567.4^{+1444.8}_{-560.7}$ & $13.75^{+0.60}_{-0.60}$ & $1.19^{+0.19}_{-0.19}$ & 56 & $20.03$ & 30.54&0.15&$0.67^*$& AJ.\\[0.5pc] 
 090424&  0.544&  $237.1^{+5.9}_{-5.9}$ & $3.54^{+0.35}_{-0.35}$ & $1.47^{+0.04}_{-0.04}$ & 52 & $18.33$ & 30.15&0.08&0.50$~^{d)}$& AK.\\[0.5pc]
 090516A&  4.11&  $725.9^{+135.2}_{-135.2}$ & $73.95^{+4.93}_{-4.93}$ & $4.69^{+0.59}_{-0.59}$ & 10.3 & $20.88^{\blacktriangledown}$ & $30.24^\blacktriangledown$&0.17&$0.84^*$& AL.\\[0.5pc]  
 090519&  3.85&  $7848.2^{+4282.6}_{-4282.6}$ & $34.30^{+2.86}_{-2.86}$ & $7.86^{+2.62}_{-2.62}$ & 74.2 & $23.14$ & 29.77&0.13&0.01$~^{g)}$& AM.\\[0.5pc]  
 090618&  0.54&  $226.2^{+5.6}_{-5.6}$ & $22.95^{+0.22}_{-0.22}$ & $1.02^{+0.02}_{-0.02}$ & 105 & $16.65$ & 30.83&0.28&0.25$~^{a)}$& AN.\\[0.5pc]   
 090812&  2.452&  $2022.9^{+838.8}_{-524.7}$ & $43.26^{+1.49}_{-1.49}$ & $8.27^{+0.46}_{-0.46}$ & 66.7 & $21.15$ & 30.18&0.08&0.41$~^{g)}$& AO.\\[0.5pc]  
 091020&  1.71&  $661.9^{+99.5}_{-99.5}$ & $6.81^{+0.18}_{-0.18}$ & $1.65^{+0.12}_{-0.12}$ & 34.6 & $18.31$ & 31.01&0.06&$0.36^*$& AP.\\[0.5pc] 
 091029&  2.752&  $229.7^{+33.8}_{-94.5}$ & $9.46^{+0.39}_{-0.39}$ & $1.71^{+0.10}_{-0.10}$ & 39.2 & $17.85$ & 31.44&0.06&$\sim 0$$~^{g)}$& AQ.\\[0.5pc]
 091127&  0.49&  $52.9^{+2.3}_{-2.3}$ & $1.39^{+0.05}_{-0.05}$ & $0.65^{+0.04}_{-0.04}$ & 7.1 & $16.77$ & 30.60&0.13&0.11$~^{a)}$& AR.\\[0.5pc]     
 091208B&  1.063&  $255.4^{+41.4}_{-40.0}$ & $2.50^{+0.15}_{-0.15}$ & $1.67^{+0.11}_{-0.11}$ & 14.9 & $19.59$ & 30.24&0.18&0.40$~^{a)}$& AS.\\[0.5pc]     
 100728B&  2.106&  $341.2^{+68.5}_{-68.5}$ & $5.66^{+0.33}_{-0.33}$ & $2.90^{+0.41}_{-0.41}$ & 10.2 & $20.11$ & 30.53&0.22&$0.35^*$& AT.\\[0.5pc] 
 100814A&  1.44&  $330.9^{+25.5}_{-25.5}$ & $11.65^{+0.26}_{-0.26}$ & $0.72^{+0.06}_{-0.06}$ & 110 & $19.02$ & 30.59&0.07&$0.11^*$& AU.\\[0.5pc] 
 110205A&  2.22&  $714.8^{+238.3}_{-238.3}$ & $57.82^{+5.78}_{-5.78}$ & $2.83^{+0.16}_{-0.16}$ & 257 & $17.33$ &  31.49&0.05&0.20$~^{a)}$& AV.\\[0.5pc] 
 110213A&  1.46&  $223.9^{+76.3}_{-64.0}$ & $8.57^{+0.58}_{-0.58}$ & $0.37^{+0.14}_{-0.14}$ & 48 & $16.02$ &  32.14&1.06&$0.07^*$& AW.\\[0.5pc]   
 110422A&  1.77&  $421.0^{+13.9}_{-13.9}$ & $66.04^{+1.61}_{-1.61}$ & $12.06^{+0.39}_{-0.39}$ & 40 & $19.04$ &  30.89&0.09&$0.65^*$& AX.\\[0.5pc]     
 110503A&  1.613&  $572.3^{+52.3}_{-49.3}$ & $17.84^{+0.71}_{-0.71}$ & $0.53^{+0.02}_{-0.02}$ & 10 & $18.89$ &  30.68&0.08&$0.15^*$& AY.\\[0.5pc]     
 110731A&  2.83&  $1223.0^{+73.4}_{-73.4}$ & $39.84^{+0.66}_{-0.66}$ & $25.49^{+0.70}_{-0.70}$ & 6.6 & $20.36$ &  30.72&0.57&0.24$~^{a)}$& AZ.\\[0.5pc]     
 120326A&  1.798&  $122.9^{+10.8}_{-10.8}$ & $4.08^{+0.47}_{-0.47}$ & $1.28^{+0.06}_{-0.06}$ & 69.6 & $18.91$ &  30.90&0.17&$0.23^*$& BA.\\[0.5pc]     
 120811C&  2.671&  $204.0^{+19.6}_{-19.6}$ & $7.42^{+0.74}_{-0.74}$ & $2.44^{+0.12}_{-0.12}$ & 26.8 & $19.42$ &  30.94&0.11&$0.53^*$& BB.\\[0.5pc] 
 120907A&  0.97&  $241.2^{+67.3}_{-67.3}$ & $0.27^{+0.04}_{-0.04}$ & $0.19^{+0.03}_{-0.03}$ & 5.8 & $20.17$ &  29.88&0.31&$0.13^*$& BC.\\[0.5pc] 
 121211A&  1.023&  $202.8^{+32.0}_{-32.0}$ & $0.49^{+0.10}_{-0.10}$ & $0.06^{+0.02}_{-0.02}$ & 182 & $19.65$ &  30.09&0.03&$0.37^*$& BD.\\[0.5pc] 
 130408A&  3.758&  $1003.9^{+138.0}_{-138.0}$ & $21.39^{+3.72}_{-3.72}$ & $21.58^{+4.40}_{-4.40}$ & 7 & $19.13$ &  31.30&0.84&$0.06^*$& BE.\\[0.5pc] 
 130420A&  1.297&  $131.6^{+7.2}_{-7.2}$ & $4.61^{+0.19}_{-0.19}$ & $0.33^{+0.02}_{-0.02}$ & 123.5 & $19.06$ &  30.51&0.04&$0.21^*$& BF.\\[0.5pc]      
 130427A&  0.3399&  $1112.1^{+6.7}_{-6.7}$ & $81.93^{+0.79}_{-0.79}$ & $11.57^{+0.16}_{-0.16}$ & 138.2 & $15.08$ &  30.96&0.07&$0.11^*$& BG.\\[0.5pc] 
 130505A&  2.27&  $2063.4^{+101.4}_{-101.4}$ & $220.26^{+10.49}_{-10.49}$ & $104.01^{+10.75}_{-10.75}$ & 88 & $17.89$ & 31.42&0.13&$0.35^*$& BH.\\[0.5pc]       
 130610A&  2.092&  $911.8^{+132.7}_{-132.7}$ & $9.59^{+0.38}_{-0.38}$ & $1.75^{+0.21}_{-0.21}$ & 46.4 & $19.88$ &  30.53&0.07&$0.23^*$& BI.\\[0.5pc]       
 130701A&  1.155&  $191.8^{+8.6}_{-8.6}$ & $3.52^{+0.08}_{-0.08}$ & $2.12^{+0.09}_{-0.09}$ & 4.4 & $19.08$ &  30.43&0.28&$0.09^*$& BJ.\\[0.5pc] 
 130831A&  0.4791&  $81.4^{+5.9}_{-5.9}$ & $0.775^{+0.002}_{-0.002}$ & $0.15^{+0.01}_{-0.01}$ & 32.5 & $17.23$ &  30.39&0.15&$0.07^*$& BK.\\[0.5pc]                   
 131030A&  1.293&  $405.9^{+22.9}_{-22.9}$ & $28.91^{+2.89}_{-2.89}$ & $5.72^{+0.14}_{-0.14}$ & 41.1 & $17.70$ &  31.04&0.19&$0.21^*$& BL.\\[0.5pc]
 131231A&  0.6419&  $291^{+6}_{-6}$ & $23.02^{+0.28}_{-0.28}$ & $2.16^{+0.02}_{-0.02}$ & 31.2 & 16.15 &  31.26 &0.07&$0.23^*$& BM.\\[0.5pc]
 140213A&  1.2076&  $191.2^{+7.8}_{-7.8}$ & $10.75^{+1.07}_{-1.07}$& $3.31^{+0.11}_{-0.11}$ & 18.6 & $18.32$ &  30.83&0.49&$0.06^*$& BN.\\[0.5pc] 
 140419A&  3.956&  $1452.1^{+416.3}_{-416.3}$ & $191.66^{+19.17}_{-19.17}$ & $28.35^{+1.16}_{-1.16}$ & 94.7 & $18.17$ &  31.67&0.10&$0.47^*$& BO.\\[0.5pc] 
 140423A&  3.26&  $532.5^{+38.3}_{-38.3}$ & $67.97^{+2.17}_{-2.17}$ & $5.93^{+0.56}_{-0.56}$ & 134 & $19.08$ &  31.12&0.04&$0.32^*$& BP.\\[0.5pc]  
 140506A&  0.889&  $373.2^{+61.5}_{-61.5}$ & $1.28^{+0.14}_{-0.14}$ & $0.74^{+0.06}_{-0.06}$ & 64.13 & $19.60$ &  30.11&0.31&$0.32^*$& BQ.\\[0.5pc] 
 140508A&  1.03&  $534^{+28}_{-28}$ & $24.53^{+0.86}_{-0.86}$ & $6.79^{+0.10}_{-0.10}$ & 44.3 & 16.78 &  31.28 &0.08&$0.20^*$& BR.\\[0.5pc]
 140512A&  0.725&  $1177.8^{+121.3}_{-121.3}$ & $9.44^{+0.20}_{-0.20}$ & $0.66^{+0.03}_{-0.03}$ & 154.8 & $18.44$ &  30.41&0.53&$0.10^*$& BS.\\[0.5pc] 
 140606B&  0.384&  $801^{+182}_{-182}$ & $0.468^{+0.04}_{-0.04}$ & $0.19^{+0.01}_{-0.01}$ & 22.8 & 19.25 &  29.65 &0.32&$0.34^*$& BT.\\[0.5pc]
 140620A&  2.04&  $387^{+34}_{-34}$ & $7.28^{+0.37}_{-0.37}$ & $3.02^{+0.17}_{-0.17}$ & 45.8 & 19.08 &  31.11 &0.15&$0.16^*$& BU.\\[0.5pc]
 140623A&  1.92&  $834^{+317}_{-317}$ & $3.58^{+0.40}_{-0.40}$ & $0.69^{+0.08}_{-0.08}$ & 114.7 & 20.07 &  30.64 &0.10&$\sim0^*$& BV.\\[0.5pc]
 140703A&  3.14&  $861.3^{+148.3}_{-148.3}$ & $20.96^{+1.61}_{-1.61}$ & $4.82^{+1.03}_{-1.03}$ &84.2 & $19.10$ &  31.28&0.34&$0.27^*$& BW.\\[0.5pc]
 140808A&  3.29&  $503^{+35}_{-35}$ & $8.71^{+0.60}_{-0.60}$ & $10.69^{+0.37}_{-0.37}$ & 4.5 & 18.23 &  32.16 &0.07&$1.1^*$& BX.\\[0.5pc]   
\enddata
\label{tab_GRB}
\end{deluxetable}

\begin{deluxetable}{ccccccc}
\tablecolumns{7}
\tablewidth{0pc}
\tablecaption{Parameters of the GRB Sample without a Redshift.}
\tablecomments{(1) GRB name where ($^\bigstar$) indicates that for this GRB a strong constraint on the redshift was determined by \cite{Volnova2014f}; (2) observed peak energy of the $\rm{\nu F_\nu}$ spectrum; (3) observed $\gamma$-ray fluence in the 15-150 keV energy band; (4) Observed 1s-peak photon flux in the 15-150 keV energy band; (5) and (6) galactic and host visual extinctions, respectively, where (*) indicates that the host extinction is estimated from the $NH_{X,i}$ measurement); (7) apparent R magnitude measured 2 hr after the burst not corrected for the galactic and host extinctions ($\blacktriangledown$ : upper limit on R mag. $\dagger$ : calibrated with the V-band magnitude).
}
\tablehead{
\colhead{GRB }& \colhead{ $E_{po}$} & \colhead{ $S_\gamma$}&\colhead{ $P_\gamma$}&\colhead{ $A_V^{gal}$}& \colhead{ $A_V^{host}$} & \colhead{ $R_{mag}$} \\
\colhead{ } &\colhead{(keV)}&\colhead{ $(\rm{10^{-7}~erg.cm^{-2}})$ }& \colhead{ $(\rm{photon.cm^{-2}.s^{-1}})$}& \colhead{ (mag)}& \colhead{ (mag)}& \colhead{ (t$_{\rm obs}$=2h)} \\
\colhead{ (1)}& \colhead{ (2)} & \colhead{ (3)} & \colhead{ (4)}& \colhead{ (5)} &\colhead{ (6)}&\colhead{ (7) }}
\startdata
051008$^\bigstar$ & $307^{+178}_{-136}$ &  $50.90^{+1.45}_{-1.45}$ & $5.44^{+0.35}_{-0.35}$ &0.04 &  0.21$^*$ & $23.30^{\blacktriangledown}$\\[0.5pc]
060105 & $327^{+40}_{-46}$ & $176.00^{+3.04}_{-3.04}$ & $7.44^{+0.36}_{-0.36}$ &  0.56 & 0.13$^*$& $20.20^{\blacktriangledown}$ \\[0.5pc]
060117 & $184^{+30}_{-30}$ & $202.00^{+3.71}_{-3.71}$ & $48.30^{+1.56}_{-1.56}$ & 0.12 &0.45$^*$& $20.15^{\blacktriangledown}$ \\[0.5pc]
060904 & $163^{+31}_{-31}$ & $77.20^{+1.51}_{-1.51}$ & $4.87^{+0.20}_{-0.20}$ &  0.06 & 0.05$^*$& $21.51^{\blacktriangledown}$ \\[0.5pc]
090813 & $95^{+30}_{-30}$ & $13^{+1}_{-1}$ &$8.5^{+0.6}_{-0.6}$ & 0.56 &0.12$^*$& $20.21^{\blacktriangledown}$ \\[0.5pc]
091221 & $207^{+22}_{-17}$ & $57^{+2}_{-2}$ & $3.0^{+0.2}_{-0.2}$ & 0.70 &0.07$^*$& 21.44 \\[0.5pc]
100413A & $446^{+123}_{-123}$ & $62^{+2}_{-2}$ & $0.7^{+0.1}_{-0.1}$ &  0.37 & 0.15$^*$& 23.20 \\[0.5pc]
101011A & $296.6^{+49.4}_{-49.4}$ & $14^{+1}_{-1}$ &$1.3^{+0.2}_{-0.2}$ & 0.10 &0.11$^*$& $20.80^{\blacktriangledown\dagger}$ \\[0.5pc]
140102A & $186^{+5}_{-5}$ & $77^{+2}_{-2}$ & $29.8^{+0.6}_{-0.6}$ & 0.11 &0.10$^*$& 19.34 \\[0.5pc]
140626A & $44.7^{+9.5}_{-9.5}$&  $3.6^{+0.5}_{-0.5}$ & $0.7^{+0.1}_{-0.1}$ & 0.47 &0.07$^*$&$20.84^{\blacktriangledown}$\\[0.5pc]
140709B & $530^{+232}_{-232}$ &  $42^{+2}_{-2}$ &$0.9^{+0.1}_{-0.1}$ &  0.13 &0.12$^*$& $20.50^{\blacktriangledown}$\\[0.5pc]
140713A & $96^{+24}_{-24}$ & $3.7^{+0.3}_{-0.3}$ & $1.9^{+0.2}_{-0.2}$ &  0.16 &0.27$^*$& $23.90^{\blacktriangledown}$\\[0.5pc]
141005A & $119^{+10}_{-10}$ & $11^{+1}_{-1}$ &$4.2^{+0.8}_{-0.8}$ & 0.56 &0.13$^*$& $21.44^{\blacktriangledown}$\\[0.5pc]
141017A & $97^{+12}_{-10}$ & $31^{+1}_{-1}$ & $6.7^{+0.3}_{-0.3}$ & 0.13 &0.12$^*$&$22.40^{\blacktriangledown}$\\
\enddata
\label{tab_GRB_sansz}
\end{deluxetable}

\newpage

\begin{deluxetable}{ll}
\tablecolumns{2}
\tablewidth{0pc}
\tablecaption{References for the afterglow optical light curve}
\tablehead{
\colhead{GRB }& \colhead{ Reference R mag} \\
}
\startdata
\cutinhead{GRBs with a Redshift}
990123 & \cite{Kulkarni1999}\\
990510 & \cite{Harrison1999}\\
990712 &\cite{Sahu2000}\\
020124 & \cite{Berger2002} \cite{Hjorth2003}\\    
020813 & \cite{Urata2003}, \cite{Laursen2003}, \cite{Li2002a} ,\cite{Li2002b}, \cite{Williams2002}\\ 
       & \cite{Beskin2002},\cite{Kiziloglu2002}, \cite{Fiore2002}, \cite{Gorosabel2002}\\ 
021004 &\cite{Pandey2003a},  \cite{Holland2003},\cite{Bersier2003}\\
       & \cite{deUgartePostigo2005}, \cite{Garnavich2002}\\    
021211 & \cite{Fox2003}, \cite{Li2003}, \cite{Holland2004}\\
       & \cite{Pandey2003b}, \cite{Testa2003}, \cite{Price2002a}\\
030328 & \cite{Maiorano2006}, \cite{Burenin2003}, \cite{Rumyantsev2003}\\
       & \cite{Fugazza2003}, \cite{Ibrahimov2003}, \cite{Martini2003}\\
030329 &\cite{Gorosabel2006}, \cite{Lipkin2004}\\
040924 &\cite{Huang2005}, \cite{Soderberg2006},  \cite{Wiersema2008}\\  
041006 &\cite{Stanek2005}, \cite{Urata2007}, \cite{Misra2005}\\ 
050416A &\cite{Kann2010}\\
050525A &\cite{Resmi2012}, \cite{Klotz2005}, \cite{Rykoff2005}, \cite{Torii2005}\\
        & \cite{Malesani2005}, \cite{Mirabal2005},  \cite{Cobb2005}, \cite{Durig2005}\\
        & \cite{Haislip2005}, \cite{Yanagisawa2005},  \cite{Kann2010}\\
050820A &\cite{Cenko2006a}, \cite{Kann2010}\\
050922C &\cite{Kann2010}\\
060115 & \cite{Yanagisawa2006}, \cite{Distefano2006}, \cite{Nysewander2006a}\\
060908 &\cite{Covino2010}\\
061007 &\cite{Mundell2007}\\
061121 &\cite{Kann2010}, \cite{Halpern2006a}, \cite{Halpern2006b}, \cite{Halpern2006c}\\
       & \cite{Yost2006}, \cite{Melandri2006},  \cite{Uemura2006}\\
070612A & \cite{Updike2007a}, \cite{Updike2007b}, \cite{Updike2007c},  \cite{Mirabal2007}, \cite{Malesani2007}\\
071010B & \cite{Lee2010}, \cite{Oksanen2008}, \cite{Wang2008}, \cite{Kann2007}, \cite{Kocevski2007}\\
071020 &\cite{Kann2010}, \cite{Schaefer2007}, \cite{Ishimura2007}, \cite{Hentunen2007}, \cite{Im2007}\\
080319B & \cite{Bloom2009}, \cite{Wozniak2009}\\
080411 & \cite{Kruehler2008a}, \cite{Thoene2008a}\\
080413A & \cite{Kann2010}, \cite{Rykoff2008}, \cite{Gomboc2008a}, \cite{Cobb2008a}, \cite{Klotz2008}\\
080413B & \cite{Brennan2008}, \cite{Gomboc2008b}\\
080605 & \cite{Zafar2012}, \cite{Kuvshinov2008}, \cite{Rumyantsev2008}, \cite{Gomboc2008c}\\
       & \cite{Kann2008}, \cite{Jakobsson2008a}, \cite{Yoshida2008a}\\
080721 & \cite{Starling2009}, \cite{Chen2008}, \cite{Huang2008}\\
080804 & \cite{Rujopakarn2008}, \cite{Kruehler2008b}, \cite{Guidorzi2008}\\
080810 & \cite{Page2009}, \cite{Kann2010}, \cite{Yoshida2008b}\\
081007 &\cite{Jin2013}\\
081008 & \cite{Yuan2010}, \cite{Cobb2008b}, \cite{Kann2010}, \cite{Cucchiara2008b}\\
081121 & \cite{Yuan2008}, \cite{Loew2008}, \cite{Cobb2008c}\\
081222 & \cite{Covino2013}, \cite{Cwiok2008},  \cite{Sonoda2008}\\
       & \cite{Kuroda2008}, \cite{Melandri2008}, \cite{Roy2008}\\
090102 & \cite{Gendre2010}\\
090418A& \cite{Yuan2009a}, \cite{Henden2009a}, \cite{Chornock2009a}\\
       & \cite{Bikmaev2009}, \cite{Pavlenko2009}\\ 
090424 & \cite{Kann2010},  \cite{Jin2013}, \cite{Urata2009},  \cite{Yuan2009b},  \cite{Mao2009}\\
       & \cite{Roy2009}, \cite{Rumyantsev2009a}, \cite{Oksanen2009}\\
       & \cite{Gorosabel2009a}, \cite{Olivares2009}, \cite{Im2009a}\\ 
090516A & \cite{Guidorzi2009}, \cite{Christie2009a}, \cite{Vaalsta2009a}, \cite{Gorosabel2009b}\\
        & \cite{deUgartePostigo2009a}, \cite{Rossi2009b}\\
090519 & \cite{Klotz2009a}, \cite{Klotz2009b}, \cite{Jelinek2009}, \cite{Rossi2009a}\\
              & \cite{Thoene2009a}, \cite{Thoene2009b}\\
090618 &\cite{Rujopakarn2009}, \cite{Li2009}, \cite{Cenko2009b}\\
       &\cite{Im2009b}, \cite{Updike2009a}, \cite{Melandri2009}, \cite{Klunko2009}\\
       &\cite{Anumapa2009}, \cite{Rumyantsev2009b}, \cite{Cano2009a}, \cite{Fernandez-soto2009}\\
       & \cite{Fatkhullin2009}, \cite{Galeev2009}, \cite{Khamitov2009}\\
090812 & \cite{Wren2009}, \cite{Smith2009a}, \cite{Haislip2009a}, \cite{Updike2009b}\\
091020 & \cite{Gorbovskoy2009}, \cite{Kann2009a}, \cite{Kann2009b}, \cite{Xu2009a} \\
       & \cite{Perley2009a}, \cite{Perley2009b}, \cite{Xin2009a}\\
091029 & \cite{Filgas2009a}, \cite{Chornock2009c}, \cite{Christie2009b}, \cite{deUgartePostigo2009c} \\
091127 & \cite{Troja2012}, \cite{Filgas2011}, \cite{Xu2009b}, \cite{Klotz2009c}, \cite{Cobb2010} \\
       & \cite{Kinugasa2009a}, \cite{Vaalsta2009b}, \cite{Andreev2009a}, \cite{Haislip2009b}\\
       & \cite{Haislip2009c}, \cite{Thoene2009c}\\
091208B & \cite{Nakajima2009}, \cite{Cano2009b}, \cite{Yoshida2009}, \cite{Xin2009b}\\
        & \cite{Kinugasa2009b},  \cite{Andreev2009b}, \cite{Xu2009c}, \cite{Updike2009c}\\
100728B & \cite{Perley2010a}, \cite{Perley2010b}, \cite{Elenin2010},\cite{Decia2010}, \cite{Olivares2010} \\
100814A & \cite{Nardini2014}\\
110205A & \cite{Zheng2012},\cite{Gendre2012}, \cite{Cucchiara2011b}\\
110213A & \cite{Cucchiara2011b}, \cite{Rujopakarn2011}, \cite{Wren2011}, \cite{Ukwatta2011}\\
        & \cite{Hentunen2011a}, \cite{Kuroda2011a}, \cite{Volnova2011}\\
        & \cite{Zhao2011}, \cite{Hentunen2011b}\\
110422A & \cite{Elunko2011}, \cite{Moskvitin2011a}, \cite{Rumyantsev2011a}\\
               & \cite{Hentunen2011c}, \cite{Xu2011a},  \cite{Melandri2011}\\
        & \cite{Jeon2011}, \cite{Kuroda2011b}, \cite{Xu2011b}, \cite{Rumyantsev2011b}\\
110503A & \cite{Klotz2011}, \cite{Kann2011}, \cite{Leloudas2011}, \cite{Tasselli2011}, \cite{Broens2011} \\
        & \cite{Davanzo2011}, \cite{Pavlenko2011}, \cite{Im2011},  \cite{Updike2011}\\
110731A & \cite{Bersier2011}, \cite{Kuroda2011c}, \cite{Moskvitin2011b}, \cite{Malesani2011b}\\
        & \cite{Tanvir2011}, \cite{Ackermann2013}\\
120326A & \cite{Melandri2014}, \cite{Klotz2012}, \cite{Hentunen2012},  \cite{Jang2012}\\
        & \cite{Zhao2012},  \cite{Xin2012a}, (254) \cite{Xin2012b}, \cite{Soulier2012}\\
        & \cite{Quadri2012},  \cite{Sahu2012}\\
120811C & \cite{Elenin2012},  \cite{Litvinenko2012}, \cite{Galeev2012a},  \cite{Galeev2012b}\\
120907A & \cite{Nevski2012} \\
121211A & \cite{Japelj2012},\cite{Kuroda2012},  \cite{Butler2012a}, \cite{Butler2012b}\\
130408A & \cite{Melandri2013a},\cite{Sudilovsky2013}, \cite{Trotter2013a},  \cite{Dereli2013} \\
130420A & \cite{Guver2013},\cite{Elenin2013a}, \cite{Guidorzi2013a},  \cite{Trotter2013b},  \cite{Hentunen2013a}\\
        & \cite{Butler2013a}, \cite{Watson2013a},  \cite{Butler2013b},  \cite{Zhao2013}\\
130427A & \cite{Vestrand2014}, \cite{Maselli2014}\\
130505A & \cite{Hentunen2013b},  \cite{Kuroda2013}, \cite{Xu2013a}\\
        & \cite{Xin2013a},  \cite{Kann2013}, \cite{Krugly2013},\cite{Watson2013b}\\
130610A & \cite{Klotz2013},  \cite{Melandri2013b}, \cite{Hentunen2013c}\\
        & \cite{Elenin2013b}, \cite{Cenko2013a},  \cite{Im2013a},  \cite{Rumyantsev2013}\\
130701A & \cite{Leloudas2013}, \cite{Cenko2013b}, \cite{Xu2013b}\\
130831A & \cite{Cano2014a}, \cite{Yoshii2013}, \cite{Guidorzi2013b},  \cite{Volnova2013a}, \cite{Xu2013c}\\
        & \cite{Izzo2013}, \cite{Xin2013b}, \cite{Sonbas2013}, \cite{Volnova2013b}\\
        & \cite{Volnova2013c}, \cite{Hentunen2013d}, \cite{Leonini2013}, \cite{Khorunzhev2013}\\
        & \cite{Masi2013}, \cite{Butler2013c}, \cite{Watson2013c}, \cite{Lee2013}\\
131030A & \cite{Moskvitin2013},  \cite{Xu2013d},  \cite{Virgili2013},  \cite{Gorbovskoy2013}, \cite{Terron2013}\\
        & \cite{Im2013b},  \cite{Perley2013}, \cite{Littlejohns2013a}\\
        & \cite{Littlejohns2013b},  \cite{Littlejohns2013c}, \cite{Littlejohns2013d}, \cite{Littlejohns2013e}\\
        & \cite{Hentunen2013e}, \cite{Tanigawa2013},  \cite{Xin2013c}, \cite{Pandey2013}\\
131231A & \cite{Singer2015}\\
140213A & \cite{Trotter2014}, \cite{Elliott2014}\\ 
140419A & \cite{Guver2014},  \cite{Zheng2014}, \cite{Butler2014a}, \cite{Littlejohns2014a} \\
        & \cite{Littlejohns2014b}, \cite{Cenko2014a}, \cite{Hentunen2014}, \cite{Choi2014b}\\
        & \cite{Kuroda2014a}, \cite{Kuroda2014b}, \cite{Volnova2014a}, \cite{Pandey2014b}, \cite{Xu2014b}\\
140423A & \cite{Ferrante2014}, \cite{Elenin2014}, \cite{Cenko2014b}, \cite{Harbeck2014a}\\
               & \cite{Harbeck2014b}, \cite{Akitaya2014},  \cite{Kuroda2014c}\\
        & \cite{Fujiwara2014}, \cite{Volnova2014b},  \cite{Sahu2014}\\
        & \cite{Cano2014b},  \cite{Volnova2014c}, \cite{Bikmaev2014}\\
        & \cite{Littlejohns2014c},  \cite{Butler2014b},  \cite{Volnova2014d}\\
140506A & \cite{Fynbo2014}\\
140508A & \cite{Singer2015}\\
140512A & \cite{Gorbovskoy2014}, \cite{Klunko2014}, \cite{deUgartePostigo2014a},  \cite{Graham2014}\\
140606B & \cite{Singer2015}\\
140620A & \cite{Singer2015}\\
140623A & \cite{Singer2015}\\
140703A & \cite{Ciabattari2014}, \cite{Perley2014b}, \cite{Butler2014c},  \cite{Volnova2014e} \\
140808A & \cite{Singer2015}\\
\cutinhead{GRBs without a Redshift}
051008 & \cite{Rumyantsev2005},  \cite{Pozanenko2005},  \cite{Volnova2014f}\\
060105 & \cite{Urata2006a}, \cite{Zimmerman2006}, \cite{Kann2006}\\
060117 & \cite{Nysewander2006b}\\
060904 & \cite{Klotz2006}, \cite{Cenko2006b}, \cite{Urata2006b} \\
090813 & \cite{Smith2009b}, \cite{Volnova2009}\\
091221 & \cite{Zheng2009}, \cite{Filgas2009b}, \cite{Haislip2009d}\\
100413A & \cite{Ivanov2010}, \cite{Guidorzi2010}, \cite{Xin2010},  \cite{Filgas2010}\\
101011A & \cite{Laas-Bourez2010}, \cite{Nardini2010}, \cite{dePasquale2010}, \cite{Tello2010}\\
140102A & \cite{Yoshii2014}, \cite{Guziy2014}, \cite{Chen2014}, \cite{Xu2014a}, \cite{Pandey2014a}\\
        & \cite{Tanga2014a}, \cite{Perley2014a}, \cite{Choi2014a}, \cite{Volnova2014g} \\
140626A & \cite{Klotz2014a}, \cite{Tanga2014b}\\
140709B & \cite{Ivanov2014}, \cite{Volnova2014i}, \cite{Volnova2014j} \\
140713A & \cite{Xin2014}, \cite{Cano2014b}\\
141005A & \cite{Perley2014c}, \cite{Schmidl2014} \\
141017A &  \cite{Klotz2014b}, \cite{Kann2014}\\
\enddata
\end{deluxetable}

\begin{deluxetable}{l}
\tablecolumns{1}
\tablewidth{0pc}
\tablecaption{References for the redshifts and Host extinctions}
\tablehead{References for the redshift}
\startdata
{{\bf A.} \cite{Kulkarni1999}, {\bf B.} \cite{Vreeswijk1999}, {\bf C.} \cite{Galama1999}, {\bf D.} \cite{Hjorth2003}}\\
{{\bf E.} \cite{Price2002b}, {\bf F.} \cite{Castro-Tirado2010}, {\bf G.} \cite{DellaValle2003}, {\bf H.} \cite{Rol2003}}\\
{{\bf I.} \cite{Thoene2007}, {\bf J.} \cite{Wiersema2008}, {\bf K.} \cite{Price2004}, {\bf L.} \cite{Cenko2005}}\\
{{\bf M.} \cite{Foley2005}, {\bf N.} \cite{Prochaska2005}, {\bf O.} \cite{Jakobsson2005},{\bf P.} \cite{Piranomonte2006}}\\
{{\bf Q.} \cite{Fynbo2009}, {\bf R.} \cite{Osip2006}, {\bf S.} \cite{Bloom2006}, {\bf T.} \cite{Cenko2007a}}\\
{{\bf U.} \cite{Cenko2007b}, {\bf V.} \cite{Jakobsson2007}, {\bf W.} \cite{Vreeswijk2008a}, {\bf X.} \cite{Thoene2008a}}\\
{{\bf Y.} \cite{Thoene2008b}, {\bf Z.} \cite{Vreeswijk2008b}, {\bf AA.} \cite{Jakobsson2008a}, {\bf AB.} \cite{Jakobsson2008b}}\\
{{\bf AC.} \cite{Cucchiara2008a}, {\bf AD.} \cite{Prochaska2008}, {\bf AE.} \cite{Berger2008a}, {\bf AF.} \cite{Cucchiara2008c}}\\
{{\bf AG.} \cite{Berger2008b}, {\bf AH.} \cite{Cucchiara2008d}, {\bf AI.} \cite{Cucchiara2009a}, {\bf AJ.} \cite{Chornock2009b}}\\
{{\bf AK.} \cite{Chornock2009d}, {\bf AL.} \cite{deUgartePostigo2009b}, {\bf AM.} \cite{Thoene2009b}}\\
{{\bf AN.} \cite{Cenko2009c}, {\bf AO.} \cite{deUgartePostigo2009d}, {\bf AP.} \cite{Xu2009a}, {\bf AQ.} \cite{Chornock2009c}}\\
{{\bf AR.} \cite{Cucchiara2009b}, {\bf AS.} \cite{Wiersema2009a}, {\bf AT.} \cite{Flores2010}, {\bf AU.} \cite{Omeara2010}}\\
{{\bf AV.} \cite{Cenko2011a}, {\bf AW.} \cite{Milne2011}, {\bf AX.} \cite{Malesani2011a}, {\bf AY.} \cite{deUgartePostigo2011a}}\\
{{\bf AZ.} \cite{Tanvir2011}, {\bf BA.} \cite{Tello2012}, {\bf BB.} \cite{Thoene2012}, {\bf BC.} \cite{SanchezRamirez2012}}\\
{{\bf BD.} \cite{Perley2012}, {\bf BE.} \cite{Hjorth2013}, {\bf BF.} \cite{deUgartepostigo2013}, {\bf BG.} \cite{Flores2013}}\\
{{\bf BH.} \cite{Tanvir2013}, {\bf BI.} \cite{Smette2013}, {\bf BJ.} \cite{Xu2013b}, {\bf BK.} \cite{Cucchiara2013}}\\
{{\bf BL.} \cite{Xu2013d}, {\bf BM.} \cite{Singer2015},{\bf BN.} \cite{Schulze2014}, {\bf BO.} \cite{Tanvir2014a}}\\ 
{{\bf BP.} \cite{Tanvir2014b},{\bf BQ.} \cite{Fynbo2014}, {\bf BR.} \cite{Singer2015},{\bf BS.} \cite{deUgartePostigo2014b}}\\
{{\bf BT./BU./BV.} \cite{Singer2015}, {\bf BW.} \cite{CastroTirado2014}, {\bf BX.} \cite{Singer2015}}\\
\cutinhead{References for the Host extinction}
{a) \cite{Japelj2014}, b) \cite{Christensen2004}, c) \cite{Schady2007a}, d) \cite{Kann2010}}\\
{e) \cite{Schady2010}, f) \cite{Schady2007b}, g) \cite{Greiner2011}}\\
\enddata
\label{tab_refz}
\end{deluxetable}

\appendix
\section{The case of GRB090519 and the need for a ground-based observational strategy}

GRB090519 is an exception as it is a very faint GRB with a redshift of $z=3.85$. It was detected by both {\it Fermi} and {\it Swift} on 2009 May 19 at 21:08:56 UT (see \cite{Perri2009}). {\it Swift}-XRT and {\it Swift}-UVOT rapidly observed the field of GRB090519 ($<$130s after the burst). An X-ray counterpart was clearly identified by {\it Swift}-XRT, allowing for a refined localization of the burst at R.A.(J2000) = $09^h29^m06.85^s$ and decl(J2000) = $+00^d10'48.6"$. However, no significant optical counterpart was detected by {\it Swift}-UVOT and a corresponding 3$\sigma$ upper limit of 19.6 mag (white filter) was estimated $\sim200$s after the burst. On the ground, fast robotic telescopes rapidly responded to the GCN notice, such as TAROT at Calern observatory in France (\cite{Klotz2009a}), FARO at Chante Perdrix Observatory in France (\cite{Klotz2009b}), and BOOTES-1B in Spain (\cite{Jelinek2009}). No optical counterpart was detected and a limiting magnitude of $R>18.5$ at $t\sim230$s after the burst was estimated based on the TAROT Calern observations. The low galactic extinction ($A_V^{Gal}=0.13$) suggested that GRB090519 is a high-z GRB or was embedded in a very dusty environment. In the next hours, NOT (see \cite{Thoene2009a}), and the GROND telescope (see \cite{Rossi2009a}), detected a new fading optical source in the XRT field of view which was associated with the afterglow emission from GRB090519. The magnitude measured by NOT at t$\sim 0.33$hr after the burst was R$\sim22.8$, reavealing that the afterglow of GRB090519 was among the faintest afterglows ever observed (figure \ref{fig_afterglow}). The faintness of the afterglow was due to a combination of a high-redshift value and an intrinsic weak luminosity of the afterglow (see figure \ref{z_vs_LR}). \cite{Greiner2011} found $AV_{Host}\sim0.01$ by fitting the broadband afterglow spectrum built form GROND and {\it Swift}-XRT data. We therefore exclude a very dusty environment surrounding GRB 090519. Although faint, the redshift of GRB090519 was determined by the VLT (see \cite{Thoene2009b}) thanks to a fast response to the GCN notice. We can reasonably assume that a delay of a few additionnal hours in the VLT observation would have made the afterglow of GRB090519 unreachable for a redshift measurement and useless for GRBs studies.

This exceptional case is a good example which demonstrates the need for an observational strategy designed to quickly determine the redshift of such under-luminous GRBs. Suc a strategy will help us to better understand the optical selection effects and reduce their impact on GRB rest-frame studies by extending the observed population of GRBs with a redshift toward the least luminous ones.

\smallskip

Consequently, the designs of the future telescopes devoted to GRBs must include large apertures (above one meter), rapid slewing (less than one minute), and the ability to quickly perform low-resolution (to estimate redshifts at +/- 0.2) and near infrared spectrometry (to detect GRBs at redshifts higher than 9).
According to that requirement, the next GRB mission SVOM will optimize the GRB detection mostly toward the antisolar direction in order to detect more easily optical afterglows from the ground. One-meter class telescopes, as previously described and largely spread over the world, will be important to avoid biases against GRBs that  are too faint for the current observatories or not observable because of local observational constraints. 


\section*{Acknowledgments}
We gratefully acknowledge financial support from the OCEVU LabEx, France. 
We are also grateful to Yves Zolnierowski for insightful discussions and to the anonymous referee for comments that led us to deepen the interpretation of the results.


\end{document}